\documentclass[preprint,5p,times,twocolumn]{elsarticle}

\usepackage{lipsum,etoolbox}

\patchcmd{\maketitle}
  {\finalMaketitle\tableofcontents\vspace{\baselineskip}}
  {}{}

\usepackage{framed} 
\usepackage{multicol} 
\usepackage{nomencl} 
\makenomenclature
\usepackage{etoolbox}
\renewcommand\nomgroup[1]{%
  \item[\bfseries
  \ifstrequal{#1}{A}{Abbreviations}{%
  \ifstrequal{#1}{B}{Mathematical symbols}{}}%
]}
\usepackage{lipsum, babel}
\renewcommand*\nompreamble{\begin{multicols}{2}} 
\renewcommand*\nompostamble{\end{multicols}}
\usepackage{longtable}
\usepackage{lipsum} 
\usepackage{float}
\usepackage{graphicx}
\usepackage{tabularx}
\usepackage{multirow}
\usepackage{geometry}
\usepackage{amsmath}
\usepackage{amsfonts,amsthm,bm} 
\usepackage{color,soul}
\usepackage{amsmath}
\floatstyle{plaintop} 
\restylefloat{table}
\usepackage[table,xcdraw,dvipsnames]{xcolor}
\usepackage{rotating}
\usepackage{smartdiagram}
\usesmartdiagramlibrary{additions}
\usepackage{epstopdf}
\usepackage{caption}
\usepackage{subcaption}
\biboptions{numbers,compress}
\usepackage{marginnote}
\usepackage{hyperref}

\usepackage{optidef}
\usepackage{mathtools}
\DeclareMathSymbol{\sm}{\mathbin}{AMSa}{"39} 
\hypersetup{
    colorlinks=true, 
    linktoc=all,     
    linkcolor=blue,  
    citecolor=blue,  
}
\usepackage{tikz}
\usetikzlibrary{mindmap}
\usetikzlibrary{trees}
\usetikzlibrary{shadings}
\tikzstyle{every node}=[draw=black,thin,anchor=west, minimum height=2.5em]

\renewcommand\paragraph{\@startsection{paragraph}{4}{\z@}%
            {-2.5ex\@plus -1ex \@minus -.25ex}%
            {1.25ex \@plus .25ex}%
            {\normalfont\normalsize\bfseries}}
\long\def\symbolfootnote[#1]#2{\begingroup%
\def\thefootnote{\fnsymbol{footnote}}\footnote[#1]{#2}\endgroup}
\def\checkmark{\tikz\fill[scale=0.4](0,.35) -- (.25,0) -- (1,.7) -- (.25,.15) -- cycle;}
\usepackage{sectsty} 
\sectionfont{\fontsize{12}{14}\selectfont}

\definecolor{main}{RGB}{11,79,108}
\definecolor{c1}{RGB}{252,213,129}
\definecolor{c2}{RGB}{255,155,113}
\definecolor{c3}{RGB}{234,218,162}
\definecolor{c4}{RGB}{67,129,193}
\definecolor{c5}{RGB}{134,222,183}
\definecolor{c6}{RGB}{184,140,167}
\definecolor{c7}{RGB}{102,137,161}
\definecolor{c8}{RGB}{132,230,248}
\definecolor{c9}{RGB}{200,159,156}
\definecolor{c10}{RGB}{138,162,158}
\definecolor{c11}{RGB}{195,190,247}
\newcommand{\nox}{$\mathrm{NO_{x}}$}

\newcommand{\ind}[1]{{_{\mathrm{#1}}}} 

\graphicspath{{./Images/}}

\nomenclature[A]{ML}{Machine Learning}
\nomenclature[A]{MPC}{Model Predictive Control}
\nomenclature[A]{LSTM}{Long-Short Term Memory}
\nomenclature[A]{ICE}{Internal Combustion Engine}
\nomenclature[A]{ECU}{Engine Control Unit}
\nomenclature[A]{NMPC}{Nonlinear Model Predictive Control}
\nomenclature[A]{SOI}{Start Of Injection}
\nomenclature[A]{FQ}{Fuel Quantity}
\nomenclature[A]{DPM}{Detailed Physical Model}
\nomenclature[A]{ESM}{Engine Simulation Model}
\nomenclature[A]{\nox}{Nitrogen Oxides}
\nomenclature[A]{RNN}{Recurrent Neural Network}
\nomenclature[A]{VGT}{Variable-Geometry Turbocharger }
\nomenclature[A]{OCP}{Optimal Control Problem}
\nomenclature[A]{SQP}{Sequential Quadratic Programming}
\nomenclature[A]{aTDC}{after Top Dead Center}
\nomenclature[A]{CAD}{Crank Angle Degree}
\nomenclature[A]{PID}{Proportional Integral Derivative}
\nomenclature[A]{HPIPM}{High-Performance Interior-Point Method}
\nomenclature[A]{QP}{Quadratic Programming}
\nomenclature[A]{BM}{Benchmark}
\nomenclature[A]{FC}{Fully Connected}
\nomenclature[A]{GA}{Genetic Algorithm}
\nomenclature[A]{RNN}{Recurrent Neural Network}
 \nomenclature[B]{$x(k)$}{state vector}
 \nomenclature[B]{$y(k)$}{output vector}
 \nomenclature[B]{$u(k)$}{input vector}
 \nomenclature[B]{$h(k)$}{hidden state}
 \nomenclature[B]{$c(k)$}{cell state}
 \nomenclature[B]{$i(k)$}{input gate}
 \nomenclature[B]{$f(k)$}{forgot gate}
 \nomenclature[B]{$g(k)$}{cell candidate}
 \nomenclature[B]{$o(k)$}{output gate}
 \nomenclature[B]{$z(k)$}{internal output of each layer}
 \nomenclature[B]{$W_{u(f,g,i,o)}$}{weight matrices to input vector $u(k)$}
 \nomenclature[B]{$W_{h(f,g,i,o)}$}{weight matrices to hidden state vector $h(k)$}
 \nomenclature[B]{$W_{FC}$}{weight matrices of fully connected layer}
 \nomenclature[B]{$T_{\text{out}}(k)$}{Engine-output torque}
 \nomenclature[B]{$P_{\text{man}}(k)$}{Intake manifold pressure }
 \nomenclature[B]{$P$}{horizon length}
 \nomenclature[B]{$w_{T_{\text{out}}}$}{MPC weight for engine-output torque}
 \nomenclature[B]{$w_{\text{NO}_x}$}{MPC weight for \nox~minimization}
 \nomenclature[B]{$w_{\text{FQ}}$}{MPC weight for fuel consumption minimization}
 \nomenclature[B]{$w_{\Delta u}$}{MPC weight for engine-output torque}
 \nomenclature[B]{$w_{s}$}{MPC weight for Constraint violation penalty}
 \nomenclature[B]{$\hat{u}(k)$}{estimated optimal control input}
 \nomenclature[B]{$\hat{y}(k)$}{estimated plant output}
 
\begin{document}

\begin{frontmatter}
\title{Deep Learning based Model Predictive Control for Compression Ignition Engines }
\author[1]{Armin Norouzi}
\ead{norouziy@ualberta.ca}
\author[1]{Saeid Shahpouri}
\author[1]{David Gordon}
\author[2]{Alexander Winkler}
\author[3]{Eugen Nuss}
\author[3]{Dirk Abel}
\author[2]{Jakob Andert}
\author[1]{\\Mahdi Shahbakhti}
\author[1]{Charles Robert Koch}
\address[1]{Mechanical Engineering Department, University of Alberta, Edmonton, Canada}
\address[2]{Teaching and Research Area Mechatronics in Mobile Propulsion, RWTH Aachen University, Germany}
\address[3]{Institute of Automatic Control, RWTH Aachen University, Germany}

\begin{abstract}
Machine learning (ML) and a nonlinear model predictive controller (NMPC) are used in this paper to minimize the emissions and fuel consumption of a compression ignition engine. In this work machine learning is applied in two methods. In the first application, ML is used to identify a model for implementation in model predictive control optimization problems. In the second application, ML is used as a replacement of the NMPC where the ML controller learns the optimal control action by imitating or mimicking the behavior of the model predictive controller. In this study, a deep recurrent neural network including long-short term memory (LSTM) layers are used to model the emissions and performance of an industrial 4.5 liter 4-cylinder Cummins diesel engine. This model is then used for model predictive controller implementation. Then, a deep learning scheme is deployed to clone the behavior of the developed controller. In the LSTM integration, a novel scheme is used by augmenting hidden and cell states of the network in an NMPC optimization problem. The developed LSTM-NMPC and the imitative NMPC are compared with the Cummins calibrated Engine Control Unit (ECU) model in an experimentally validated engine simulation platform. Results show a significant reduction in Nitrogen Oxides (\nox) emissions and a slight decrease in the injected fuel quantity while maintaining the same load. In addition, the imitative NMPC has a similar performance as the NMPC but with a two orders of magnitude reduction of the computation time. 
\end{abstract}
\end{frontmatter}


\begin{table*}[ht!]
\begin{framed}
\begin{footnotesize}
\printnomenclature
\end{footnotesize}
\end{framed}
\end{table*}

\section{Introduction}

Internal Combustion Engines (ICEs) are ubiquitous and they power small devices such as lawn mowers all the way to large ship engines \cite{mpcreviean}. The durability and reliability of ICEs has made them very attractive. However, the broad use of the ICE can be attributed to more than 20\% of the total Greenhouse Gas (GHG) emissions worldwide~\cite{2020CCTASVMDiesel, norouzi2020correlation}. Another area of concern of ICE powered vehicles is the difference between emissions measurements taken at the test bench and the emissions that are actually emitted from tailpipe of vehicles on the road. This has resulted in new emissions legislation which now focuses on the Real Driving Emissions (RDE) in addition to test bench testing. This has led to increasing challenges for ICEs to meet current and upcoming legislation~\cite{mpcreviean}. These tighter emissions and fuel economy legislation, in combination with the complexity of the combustion process, have led to the requirement for significantly more advanced engine controllers than are currently used. New progress for ECU hardware has enabled significant improvements in the computational power of micro controllers. This increase in computational power has enabled online optimization methods to further reduce ICE emissions~\cite{mpcreviean}.

Historically, feed-forward control and feedback based controllers have been used for the control of ICE's in automotive applications. The most common feed-forward controller is the use of two-dimensional look-up tables or so-called calibration maps. Where the controller is calibrated for a variety of operating points on the test bench~\cite{guzzella2009introduction, isermann2014engine}. With the addition of a feedback controller such as a Proportional Integral Derivative (PID) controller it is possible to correct for disturbances and parameter variations~\cite{guzzella2009introduction}. The gains of the PID controller are first tuned using the procedure of parameter optimization while fine tuning is completed using the trial-and-error method. To comply with fuel consumption and emissions regulations, a significant number of control inputs for a broad range of engine speed and load conditions must be evaluated. This leads to a very time-consuming and costly calibration process for the engine manufacturer.

One solution to this problem is use of model-based feedback controllers. Several model-based controllers have been used in engine feedback control over the last five decades~\cite{Powell1987ReviewPaper} including Linear Quadratic Regulator~(LQR)~\cite{lopez2011lqr}, Linear Quadratic Gaussian~(LQG) \cite{pfeiffer2004system}, Sliding Model Control~(SMC)~\cite{norouzi2019integral}, Adaptive~\cite{souder2004adaptive}, and Model Predictive Control~(MPC) \cite{ICEMPC7, ICEMPC8}. Each of these model-based controllers has its own advantages and disadvantages; however, MPC is one of the most promising control methods since it has the ability to control highly constrained nonlinear systems. MPC is ideal for ICEs since: i) ICE combustion is significantly nonlinear, ii) all system inputs are generally constrained based on physical limitations, and iii) ICE deals with multi-objective optimization including fuel economy, emissions, noise, and operator satisfaction (i.e., load following during engine transients). Additionally, MPC can provide an optimal real-time solution for meeting the multi-objective goals of time-critical systems~\cite{ICEMPC7, ICEMPC8, WINKLER2021359}.

Beginning in 1998, MPC was applied to ICE control for the first time in a simulation environment for the air-fuel ratio (AFR) control of an spark ignition (SI) gasoline engine using a neural network based linear AFR model~\cite{ICEMPC35}. Then in 1999, using a GT-Power engine model simulations, the idle speed control of an SI engine was explored using MPC~\cite{ICEMPC29}. These early works were done in simulation environments, while recent works~\cite{ICEMPC15, saerens2008model, broomhead2016economic, yin2020model, stewart2008model} include experimental implementation of nonlinear multi-objective MPC on real engines. MPC has been used in ICE control for a wide variety of applications. Among these, MPC has been used widely for control of fuel consumption~\cite{ICEMPC6,ICEMPC14,ICEMPC1,ICEMPC3,ICEMPC2}, combustion phasing~\cite{ICEMPC7, ICEMPC8,ICEMPC6,ICEMPC14}, cyclic variability~\cite{ICEMPC6}, torque and load \cite{ICEMPC8, ICEMPC6, ICEMPC14, ICEMPC3, nuss2019nonlinear}, idle speed~\cite{ICEMPC29}, intake airpath control~\cite{ICEMPC6,ICEMPC11,ICEMPC1,ICEMPC5,ICEMPC12,ICEMPC16}, knock and Maximum Pressure Rise Rate (MPRR)~\cite{ICEMPC8,ICEMPC6}, engine-out emissions~\cite{ICEMPC15,ICEMPC14,ICEMPC1,  liu2021simultaneous}, and exhaust aftertreatment~\cite{ICEMPC11}.

MPC is a promising control technique which has been gaining significant attention due to the following five main advantages: 1) implicitly considers constraints on state, input and output variables, 2) provides closed loop control performance and stability for the optimal problem with constraints, 3) exploits the use of a future horizon while optimizing the current control law, 4) offers the possibility of both offline and real-time implementations, and 5) provides the capability to handle uncertainty in system's parameters, delays, and non-linearity in the model~\cite{mpcreviean}. However, like other model based control strategies, the challenge with MPC is the requirement for an accurate plant model and the increased computational effort that is required to evaluate the model over the prediction horizon. These two aspects can work against each other as increasingly complicated models are introduced to minimize model/plant mismatch resulting in a significant increase in computation~\cite{bemporad2020model}.

To solve these challenges of MPC, the integration of ML and MPC has been offered as an emerging area that offers the potential to enhance the control performance of MPC. Combining ML and MPC first began in the early 2000s but has since accelerated significantly in the last five years. ML integration with MPC, sometimes referred to as learning MPC, is divided into three main categories of: (1) ML in modeling, (2) ML in computational reduction, and (3) MPC in safe learning~\cite{mpcreviean}. ML in modeling is used to create a high fidelity model of the plant to be used inside MPC. ML in computational reduction is aimed at reducing the high computational costs associated with the optimization in MPC. One method used to reduce the computational time of MPC is cloning the behaviour of an MPC using ML. In this case, the online MPC is replaced by an ML model of the controller~\cite{mpcreviean, hewing2020learning}. In safe learning, an  ML learning controller is used but MPC is used to ensure that constraints are satisfied and system stability can be guaranteed.

Physics-based modeling provides accurate models based on physical insight into the system, but the engine combustion and emissions models utilize detailed 3D models which are computationally expensive~\cite{shahpournyenergies, davidsvmhcci}. This makes physical models impractical for real-time model-based control. Current production ECUs are not capable of running these detailed physical models and thus these models cannot be implemented for real-time control. An alternative approach for engine and emission modeling is ML based modeling which uses measurement data to train a data-driven model. These models can be as accurate as physical models while requiring significantly less computational time. This is desired for implementation of the model-based controllers in ECUs~\cite{shahpournyenergies}. Using ML in modeling has been successfully implemented for diesel engine using support vector machine (SVM)~\cite{ICEMPC7, ICEMPC8, MECC2021Khoshbakht}, neural network~\cite{ICEMPC35, hu2018nonlinear}, and Extreme Learning Machine (ELM)~\cite{VAUGHAN201518, janakiraman2015nonlinear}. One of the methods explored in this study is using a Deep Recurrent Neural Network such as a Long-Short Term Memory (LSTM) network. 

A Recurrent Neural Network (RNN) is structurally similar to a feedforward neural network with the exception of backward connections used to handle sequential relations. The advantage of the RNN compared to conventional feedforward neural network for dynamic modeling is its computational efficiency which is the result of parameter sharing. {However, conventional RNN cannot accurately capture any long-term dependencies of the model as the prediction can be described as utilizing a ``vanishing gradient''. This results in the contribution of earlier steps becomes increasingly small.} To solve this long term memory problem of RNN, various types of cells with long-term memory have been introduced. The most popular and well-known of these long-term memory cells is LSTM~\cite{geron2019hands}. Therefore, a deep network with LSTM layer is capable of predicting the dynamics of a system with high accuracy. To the author's knowledge, a deep network with LSTM layer has not been used for emission modeling for MPC in ICE control problem to date. 

Even with ML-based modeling, MPC computational times (especially for deep networks) is high. Additionally, depending on the complexity of the control problem, convexity, and dynamic time scale it might not be feasible for real-time implementation. ICEs are complex nonlinear systems to control and therefore high computational effort is required for real-time implementation. One method to reduce the MPC computational load is to replace the MPC with an ML controller. The ML is trained to mimic the MPC controller's behavior to achieve a significant computational time reduction for real-time implementation. For the original implementation the optimization is done using a powerful prototype ECU. Using this online implementation the ML function can be trained with data collected from the online MPC output. This ML function mimics the output of the online MPC, however, it can be implemented with a significant reduction in the the computational time compared to the online MPC. Here, to reduce the computational time of the MPC, an ML based imitation of the MPC controller is proposed where a deep network is trained using the MPC inputs and outputs. Imitation MPC has been previously explored in the control of heating, ventilation, and air conditioning (HVAC) systems~\cite{drgovna2018approximate, karg2018deep}, vehicle dynamics control~\cite{sun2018fast, zhang2019safe}, robotics~\cite{cera2018multi}, and power conversion~\cite{novak2020supervised} industries and has shown great success. {The application of imitation MPC to ICE control has also explored in the literature in terms of approximate MPC~\cite{MORIYASU2019104114, MORIYASU2018542rtrtrt} and inverse model controller~\cite{peng2022neural}. However, the application of deep RNN especially using LSTM in imitation MPC to ICE control is lacking in the literature.} 

The main contributions of this paper are
\begin{enumerate}
    \item Machine Learning based modeling for MPC design:
    \begin{enumerate}
    \item Developed a transient engine performance and emission model based on Long-Short Term Memory (LSTM) that is capable of providing a high accuracy model for nonlinear model predictive control,
    \item Developed a novel approach to augment LSTM in NMPC problems (LSTM-NMPC) by augmenting LSTM hidden and cell state into a nonlinear optimization problem,
    \item Designed an NMPC based on an ML model to minimize engine-out emission and fuel consumption while maintaining the same output torque performance when compared with a benchmark model, i.e., Cummins calibrated ECU-based Engine Simulation Model (ESM).
    \item Compared \texttt{acados}, Matlab \texttt{fmincon}, and EMBOTECH \texttt{FORCES PRO} solvers and verified real-time implementation of \texttt{acados} in Processor-in-the-loop (PIL) setup
\end{enumerate}
\item Machine Learning based control to reduce the computational time of MPC:
    \begin{enumerate}
    \item Developed an imitation based controller using deep neural network to clone the behavior of LSTM-NMPC to reduce computational time of optimization with same NMPC performance 
\end{enumerate}
\end{enumerate}

\begin{figure*}[ht!]
    \centering
    \includegraphics[width = 0.85\textwidth]{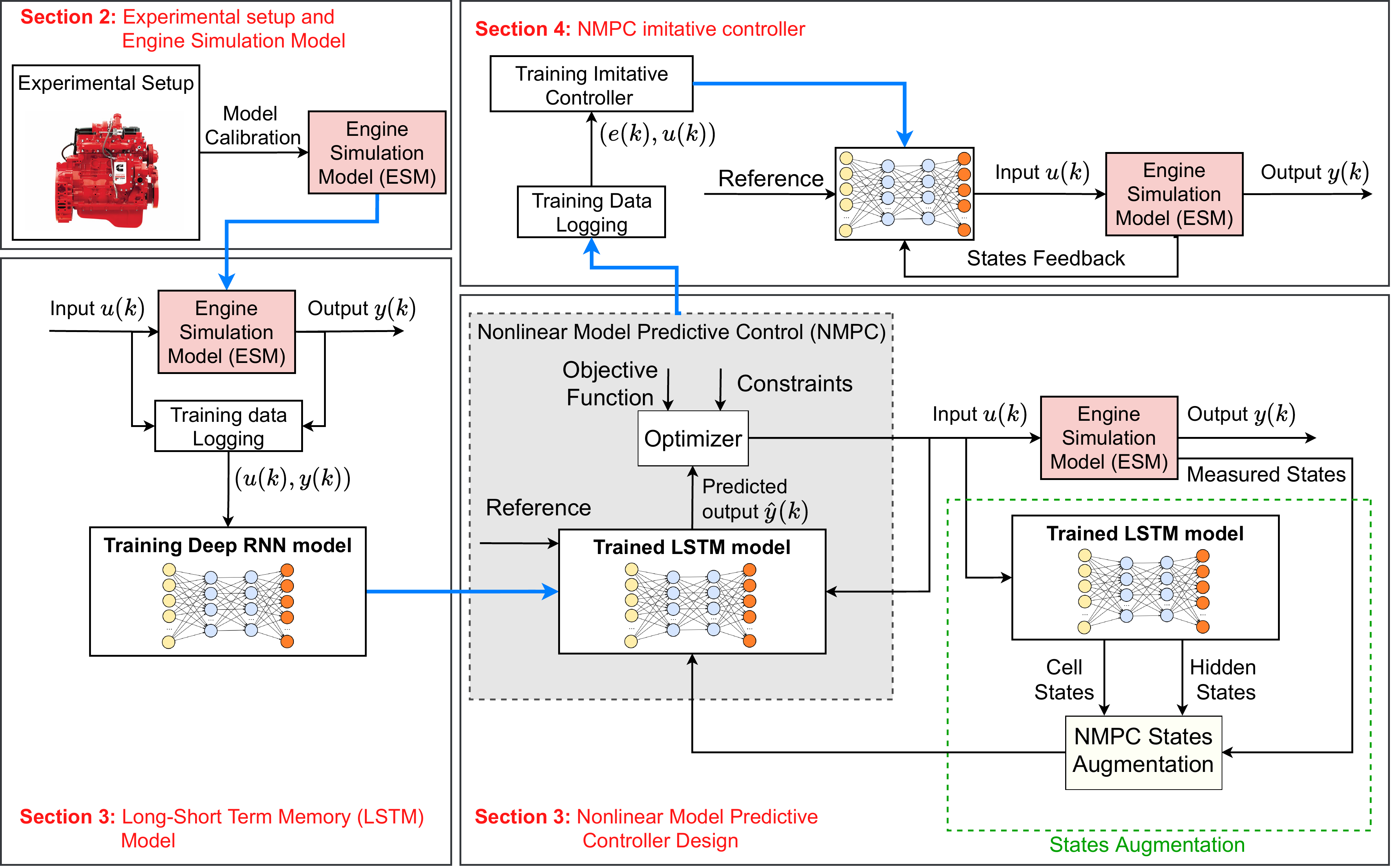}
    \caption{{Modeling and controller design procedure based on Engine Simulation Model (ESM)}}
    \label{fig:offlinemodeling}
\end{figure*}

The structure of the paper is shown in Figure~\ref{fig:offlinemodeling} and it includes logging data, data-driven model development, and controller implementation using ESM co-simulation. The first step including experimental data collection and detailed physical modeling using GT-power was done in our previous papers~\cite{saeed2021MECC, shahpournyenergies}. Then, randomly generated inputs are fed into the GT-power model and the output data is recorded for modeling. Using these input-output pairs of data, the LSTM model can then be developed. This model is used for the design of a NMPC controller. Finally, this NMPC controller is used to train the ML based imitation controllers. {All of the controllers developed in this study are simulated using an ESM co-simulation.}

\section{Engine Simulation Model (ESM)}

The engine platform used in this paper is a 4.5-liter medium-duty Cummins diesel engine. {Physical modeling of this engine is carried out using the GT Power software, which contains chemical and gas exchange sub-models that simulate the complex combustion processes. The proposed control strategies are tested in simulation using an Engine Simulation Model (ESM) parameterized to the real engine~\cite{saeed2021MECC, shahpournyenergies}. The combustion model is calibrated using experimental data where approximately 15\% of the raw data is used to calibrate the combustion model using the Genetic Algorithm (GA) algorithm in the GT-suite\(^{\copyright}\) software. The calibration process uses NSGA-III~\cite{deb2013evolutionary} for multi-objective Pareto optimization as the search algorithm.} The DIPulse model is employed as the combustion model because it can deal with multi-injection combustion engines. An extended Zeldovich \nox~model is added to the combustion model in order to add a \nox~prediction to the ESM. Additional details regarding the model structure and accuracy of the ESM can be found in~\cite{saeed2021MECC, shahpournyenergies}. This model has an accuracy of $\pm$ 5.8\%, $\pm$ 4.6\%, and $\pm$ 18.1\% in prediction of maximum in-cylinder pressure, intake manifold pressure, and \nox. This ESM is used to develop data-driven models, design controllers, and evaluate the developed controllers.



\begin{figure*}[ht!]
    \centering
    \includegraphics[width = 0.65\textwidth]{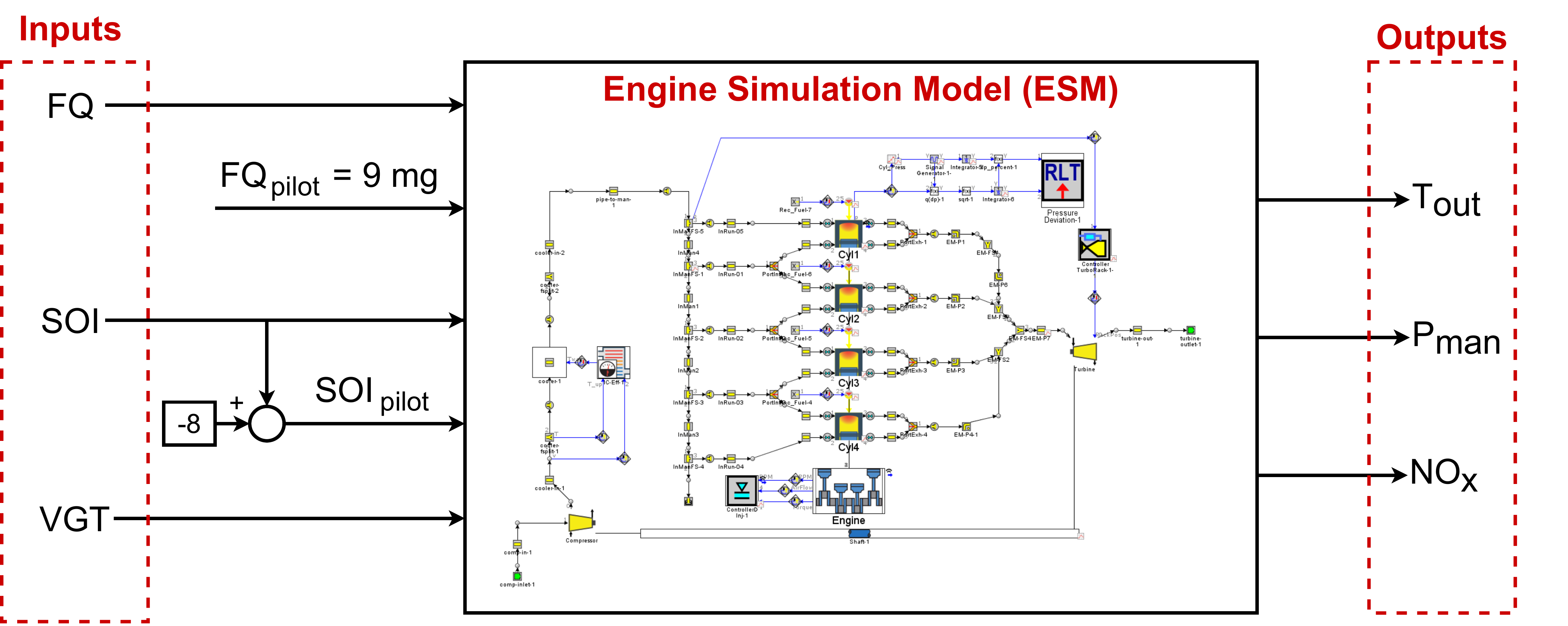}
    \caption{Schematics of Engine Simulation Model (ESM)}
    \label{fig:modelschem}
\end{figure*}

The main inputs and outputs of this ESM are shown schematically in Figure~\ref{fig:modelschem}. The inputs of this model are Fuel Quantity (FQ), Start Of Injection (SOI) for main injection, and Variable Geometry Turbine (VGT) rate. For simplicity, the pilot injection is kept constant at 9~mg of injected fuel per cycle. {The pilot injection timing varies with the main SOI, however, is kept at a constant 8 Crank Angle Degree (CAD) before the main injection.} By adding these constraints between inputs it was possible to reduce the control action from 5 manipulated variables to 3. The outputs of the ESM are output torque ($T_{\text{out}}$), intake manifold pressure $P_{\text{man}}$, and \nox~ emissions. 

\section{Long-Short Term Memory Network (LSTM) Model}

\begin{figure}[ht!]
    \centering
    \includegraphics[width = 0.49\textwidth]{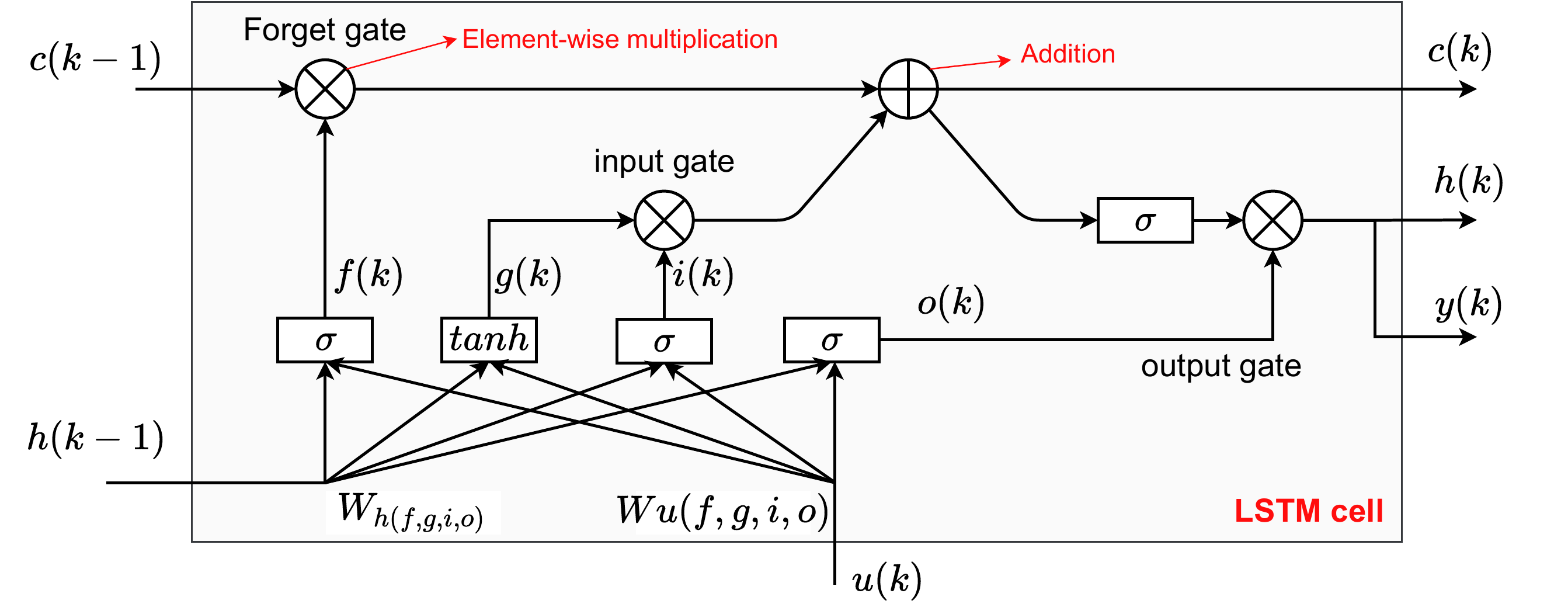}
    \caption{Long-Short Term Memory (LSTM) cell structure}
    \label{fig:LSTMCell}
\end{figure}

LSTM is one of the most popular and well-known variants of Recurrent Neural Networks (RNN). In comparison to RNN, LSTM utilizes a hidden state that is split into two main parts as shown schematically in Figure~\ref{fig:LSTMCell}: \(h{(k)}\) the short-term state, and \(c{(k)}\) the long-term state. The long-term state \(c{(k-1)}\) travels though the network and first enters the forget gate \( f(k)\) where past values are dropped. Then additional values (memories) are added to the input gate \( i(k)\) at each time step. Therefore at each time step some data is added, and some is dropped. Further, after adding new memory, the long-term state is replicated, passed into the hyperbolic tangent activation function ($tanh$), and the output gate filters the result to generate the short-term state \(h{(k)}\) (equal to the cell’s output \(y{(k)}\) for this time step)~\cite{geron2019hands}. LSTM computations can be summarized as:
\begin{equation}\label{eq:LSTMeq}
\begin{split}
i{(k)} &= \sigma\left(W_{ui}^Tu(k) + W_{hi}^T h{(k-1)} + b_i\right) \\
f{(k)} &= \sigma\left(W_{uf}^Tu(k) + W_{hf}^T h{(k-1)} + b_f\right) \\
g({k}) &=  \text{tanh} \left(W_{ug}^Tu(k) + W_{hg}^T h{(k-1)} + b_g\right)\\
o{(k)} &= \sigma\left(W_{uo}^Tu(k) + W_{ho}^T h{(k-1)} + b_o\right) \\
c{(k)} &=  f{(k)} \odot c{(k-1)} + i{(k)} \odot g({k})\\
h{(k)} &= y{(k)} = o{(k)} \odot \text{tanh} \left(c{(k)}\right)\\
\end{split}
\end{equation}
where \(W_{u(f,g,i,o)}\) and \(W_{h(f,g,i,o)}\) are the weight matrices to input vector \(u(k)\) and previous short-term state \(h(k)\). In this equation, \(\odot\), is an element-wise multiplication and \(b_{(f,g,i,o)}\) are the biases. In Eq.~\ref{eq:LSTMeq}, $i(k)$, $f(k)$, $g(k)$, $o(k)$, $c(k)$, and $h(k)$ are input gate, forgot gate, cell candidate, output gate, cell state, and hidden state, respectively. The activation functions used in LSTM computation are $tanh(z)$ and $\sigma(z)$ which are defined as
\begin{equation}\label{eq:sig}
\begin{split}
\sigma(z) & = \frac{1}{1 + e^{-z}} \\
\end{split}
\end{equation}
\begin{equation}\label{eq:tanh}
\begin{split}
\tanh(z) & = \frac{e^{2z}-1}{e^{2z}+1} \\
\end{split}
\end{equation}

\begin{figure*}
    \centering
    \includegraphics[trim = 0 0 0 0, clip, width = 0.99\textwidth]{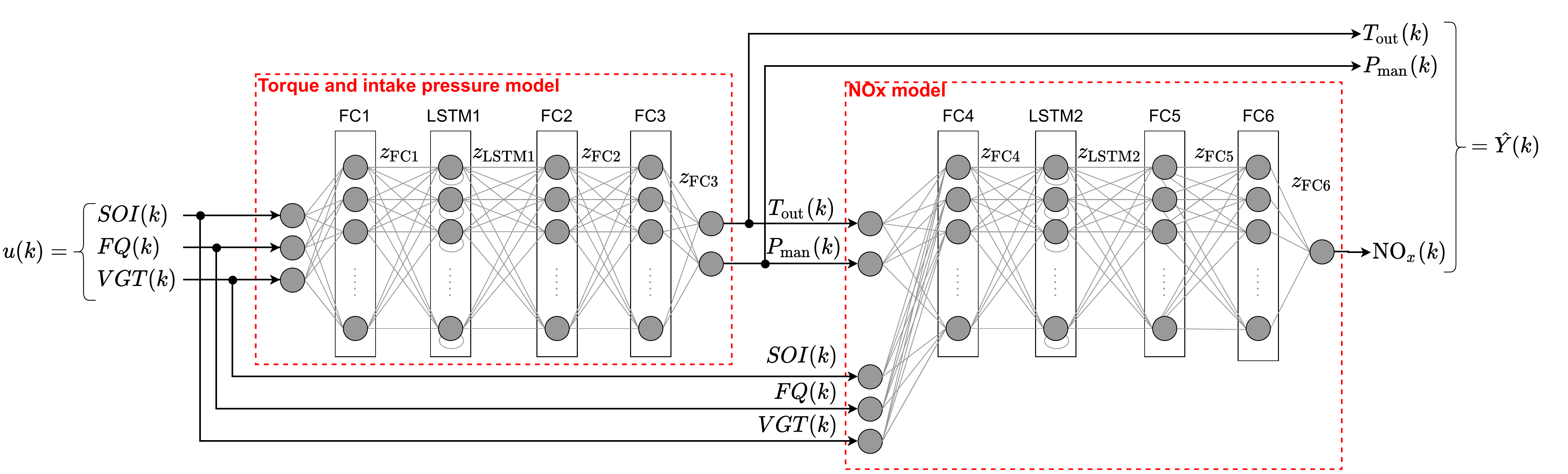}
    \caption{Structure of proposed deep neural network model for engine performance and emission modeling. FC: Fully connected layer, LSTM: Long-short term memory.}
    \label{fig:lstmmodelstructure}
\end{figure*}

A 2-level deep network, including two series networks, one for predicting intake manifold pressure and output torque and another for \nox~ emissions is proposed. Each network contains one LSTM layer and three fully connected (FC) layers. The structure of this network is shown schematically in Figure~\ref{fig:lstmmodelstructure}. The main reason for having two separate networks is a physical understanding of the system. As \nox~is created during combustion and usually measured through a sensor after each combustion cycle, it depends not only on $u(k)$ but also on intermediate states such as intake manifold pressure. For modeling \nox, having output torque is also helpful and adds more features for improving prediction accuracy. Based on a physical understanding of the system, \nox~depends on all five features, but the other two outputs, including output torque and intake manifold pressure, are not dependent on \nox. As LSTM is used in this architecture and has a recurrent behavior, adding all outputs in a single network makes intake manifold pressure and output torque a function of \nox, which is incorrect from a physical point of view.

The FC layer is added around the LSTM layer to increase the network capacity to better estimate the nonlinearity of the engine emissions and performance without increasing the number of hidden and cell states. Increasing the number of LSTM layers increases the number of hidden states which makes the subsequent NMPC problem larger. A fully connected equation with ReLU activation function is defined as 
\begin{equation}
    z_{FC}(k) = \text{ReLU}(W_{FC}^T u(k) + b_{FC})\\ \label{eq:z_FC}
\end{equation}
where \( \text{ReLU}\) is a Rectified Linear Unit (ReLU) activation function which is defined as
\begin{equation}\label{eq:relu}
\begin{split}
\text{ReLU}(z) &= \begin{cases}
         0 & \text{if } z\leq0 \\
         z & \text{if } z>0 \\
    \end{cases}
\end{split}
\end{equation}

As state and output functions are in the form of $x(k+1) = F(x(k),u(k))$ and $y(k) = F(x(k),u(k))$ need to be imported in the NMPC framework, the equation used as a predictive model of MPC is presented here. To generate state and output functions, the forward propagation needs to be evaluated for the {proposed} network in Figure~\ref{fig:lstmmodelstructure}. The forward propagation of the first part of the proposed network is used to estimate engine-out torque and intake manifold pressure are given as
\begin{subequations}
\begin{align}
\begin{split}
z_{FC1}(k) &= \text{ReLU}\left(W_{FC1}^T {u(k)} + b_{FC1}\right)\\
\end{split}\\
\begin{split}
h_{LSTM1}{(k)} &= \underbrace{o_{LSTM1}{(k)} \odot \text{tanh}\left(c_{LSTM1}{(k)}\right)}_{\text{based on Eq.}~\ref{eq:LSTMeq}}\\
\end{split}\\
\begin{split}
z_{FC2}(k) &= \text{ReLU}\left(W_{FC2}^T h_{LSTM1}{(k)} + b_{FC2} \right)\\
\end{split}\\
\begin{split}
\underbrace{z_{FC3}(k)}_{[T_{\text{out}}(k) \ \ P_{\text{man}}(k)]^T}  &= \text{ReLU}\left(W_{FC3}^T z_{FC2}(k) + b_{FC3}\right)\\
\end{split}
\end{align}
\end{subequations}


{Each of the above equations refers to the output of each layer in the left part of Figure}~\ref{fig:lstmmodelstructure} where \( u(k) \) are the system inputs which are defined as
\begin{equation}
    \begin{split}
        u(k) &= [
  \begin{array}{ccc}
    \text{FQ}(k) & \ \text{SOI}(k) & \ \text{VGT}(k)\\
  \end{array}
]^T \\
    \end{split}
\end{equation}

The estimated intake manifold pressure $P_{\text{man}}$ and output torque $T_{\text{out}}$ are then augmented with system inputs \(u(k)\) to estimate the engine-out \nox~ emissions as
\begin{subequations}
\begin{align}
\begin{split}
z_{FC4}(k) &= \text{ReLU}(W_{FC4}^T  \left[u(k) \ \ \underbrace{z_{FC3}(k)}_{[T_{\text{out}}(k) \ \ P_{\text{man}}(k)]^T} \right] + b_{FC4})\\
\end{split}\\
\begin{split}
h_{LSTM2}{(k)} &= \underbrace{o_{LSTM2}{(k)} \odot \text{tanh}\left(c_{LSTM2}{(k)}\right)}_{\text{based on Eq.}~\ref{eq:LSTMeq}}\\
\end{split}\\
\begin{split}
z_{FC5}(k) &= \text{ReLU}(W_{FC5}^T h_{LSTM2}{(k)} + b_{FC5})\\
\end{split}\\
\begin{split}
\underbrace{z_{FC6}(k)}_{\text{NO}_x(k)} & = \text{ReLU}(W_{FC6}^T z_{FC5}(k) + b_{FC6})\\
\end{split}
\end{align}
\end{subequations}


The output of this network can be calculated as
\begin{equation}
    \begin{split}
    \hat{Y}(k) &= [\underbrace{T_{\text{out}}(k) \ \ P_{\text{man}}(k)}_{z_{FC3}(k)} \ \ \underbrace{\text{NO}_x(k)}_{z_{FC6}(k)}]^T = \left[\begin{array}{c}
   z_{FC3}(k) \\
   z_{FC6}(k) \\
  \end{array} \right]
    \end{split}
\end{equation}
The augmented states of this model with hidden states and cell state to use inside NMPC are
\begin{equation} \label{eq:augmentedstates}
\begin{split}
x(k) &= \left[\begin{array}{c}
   h_{LSTM1}{(k)} \in\mathbb{R}^{26}\\
   h_{LSTM2}{(k)} \in\mathbb{R}^{26} \\
   c_{LSTM1}{(k)} \in\mathbb{R}^{26} \\
   c_{LSTM2}{(k)} \in\mathbb{R}^{26}
  \end{array} \right] \in\mathbb{R}^{104}
\end{split}
\end{equation}

\begin{figure}[ht!]
    \centering
    \includegraphics[trim = 0 0 0 0, clip, width = 0.49\textwidth]{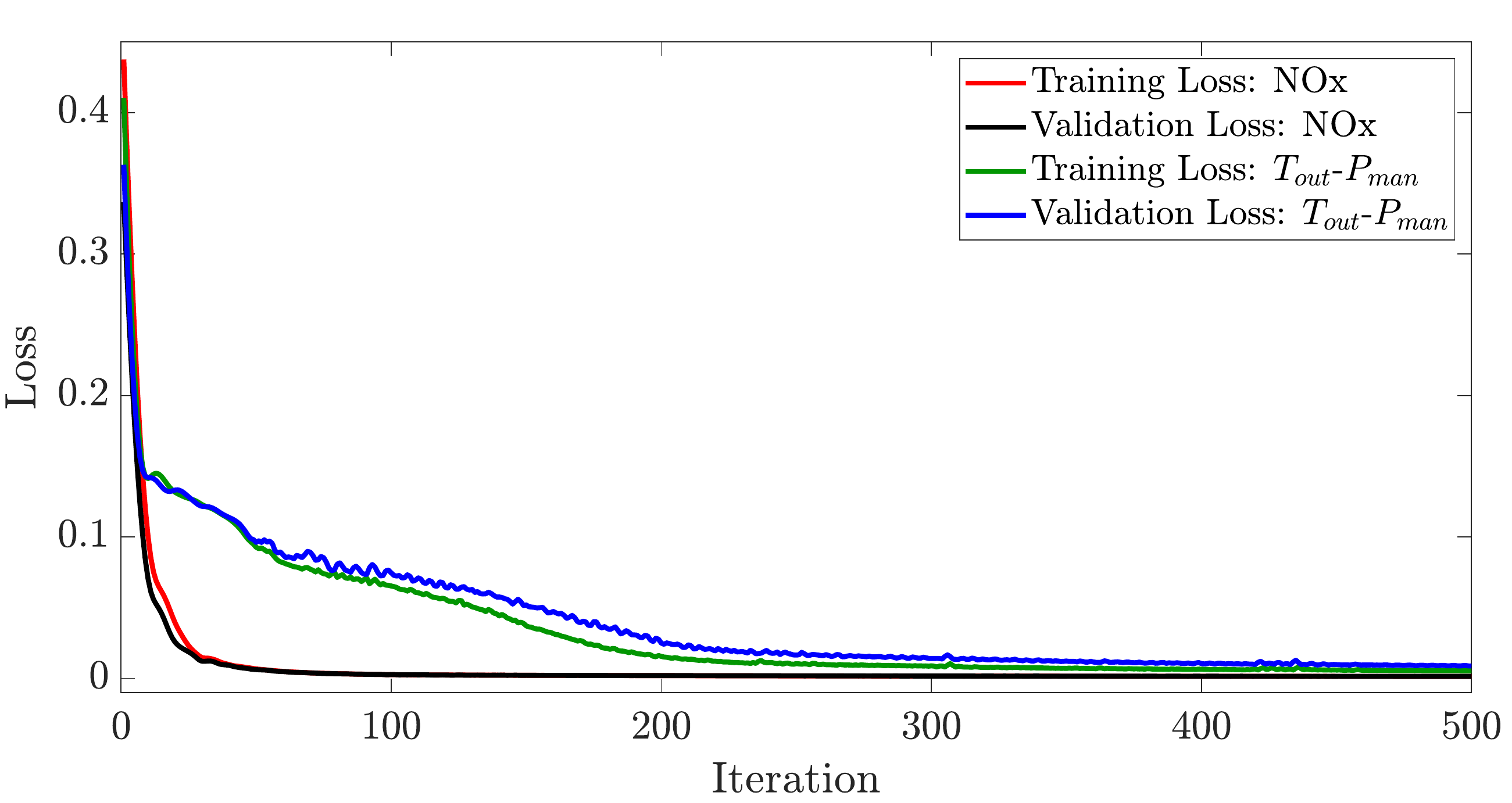}
    \caption{Loss vs. Iteration for \nox, torque and pressure model}
    \label{fig:lossfunctionlstm}
\end{figure}

The cost function for this network is as
\begin{equation}\label{eq:costreg}
J(W, b) = \frac{1}{m}  \sum_{k = 1}^m \mathcal{L}\left(\hat{Y}(k), Y(k)\right) + \frac{\lambda}{2m} \sum_{l = 1}^L ||W^{[l]}||_2^2
\end{equation}
where \(\mathcal{L}\left(\hat{Y}(k), Y(k)\right)\) is the loss function, \(\lambda\) is the regularization coefficient, and \(||W^{[l]}||_2^2\) is the Euclidean norm which is defined as
\begin{equation}\label{eq:wreg}
||W^{[l]}||_2^2 = \sum_{i = 1}^{n^{[l]}} \sum_{j = 1}^{n^{[l-1]}} (w_{ij}^{[l]})^2
\end{equation}
The Mean Squared Error (MSE) cost function is used and it is defined as
\begin{equation}
    \mathcal{L}\left(\hat{Y}(k), Y(k)\right) = \frac{1}{m} \sum_{k = 1}^m (\hat{Y}(k) - Y(k))^2
\end{equation}
The training information along with the design values for the proposed network are summarized in Table~\ref{tab:lstmmodeldetail}. To train this model, MATLAB Deep Learning Toolbox\(^{\copyright}\) utilizing the Adam algorithm has been used in this work. In Figure~\ref{fig:lossfunctionlstm}, the loss function versus iteration for both the performance and emission networks are shown, where within a defined number of Epochs, the loss functions converges to a minimum. The validation loss function also converges to match the training function, indicating that no overfitting or underfitting of the model has occurred. 

\begin{table}[ht!]
    \centering
    \caption{Properties of  2-level engine performance and emission LSTM-based model}
    \begin{tabular}{l l}
    \hline
       \textbf{Name}  &  \textbf{Value}\\
       \hline
        FC(1,2,3,4,5,6) size & 26\\
        LSTM(1,2) size & 26\\
        Optimizer & Adam\\
        Maximum Epochs & 500\\
        Mini batch size & 512\\
        Learn rate drop period & 150 Epochs\\
        Learn rate drop factor & 0.5\\
        L2 Regularization & 0.1 \\
        Initial learning rate & 0.001 \\
        Validation frequency & 1\\
        Momentum & 0.9 \\
        Squared gradient decay  & 0.99\\
        \hline
    \end{tabular}
    \label{tab:lstmmodeldetail}
\end{table}

\begin{figure}[ht!]
    \centering
    \includegraphics[trim = 10 40 50 0, clip, width = 0.49\textwidth]{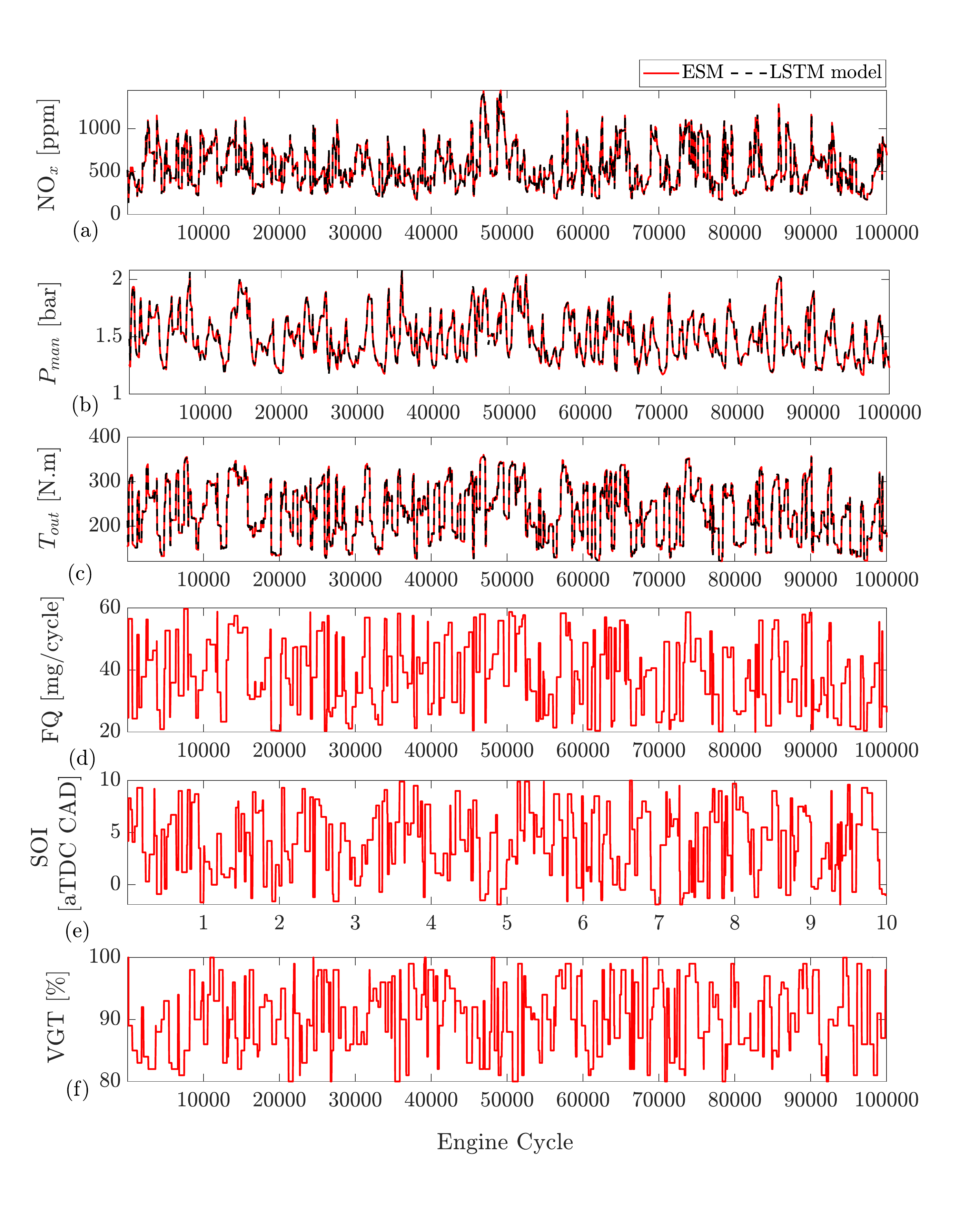}
    \caption{Training and validation results for the LSTM model vs. ESM: a) engine-out \nox, b) intake manifold pressure \( P_{\text{man}}\), c) engine-output torque \(T_{\text{out}}\), d) Fuel quantity (FQ), e) Start of injection (SOI), f) Variable Geometry Turbine ($\text{VGT}$) rate -- cycles 1 to 80,000 are devoted for training and cycles 80,001 to 100,000 used for validation}
    \label{fig:lstmtrainingvalidation}
\end{figure}

Figure~\ref{fig:lstmtrainingvalidation} shows the training and validation results of the proposed model. To develop this neural network based model, which has more than 11,000 learnable parameters, a larger data set was required. Therefore, the ESM was run for 100,000 engine cycles. Referring to Figure~\ref{fig:lstmtrainingvalidation}, cycles 1 to 80,000 are devoted for training and cycles 80,001 to 100,000 are used for validation. The LSTM model can estimate intake manifold pressure, output torque, and \nox~ emission with high accuracy for both training and validation data sets. The accuracy of the training data is 2.35\%, 1.98\%, and 1.07\% for \nox, \(T_{\text{out}}\), and \( P_{\text{man}}\), respectively. For the validation data set, an accuracy of 2.86\%, 2.27\%, and 1.53\% for \nox, \(T_{\text{out}}\), and \( P_{\text{man}}\) are found. 

\begin{figure}[ht!]
    \centering
    \includegraphics[trim = 10 40 40 0, clip, width = 0.49\textwidth]{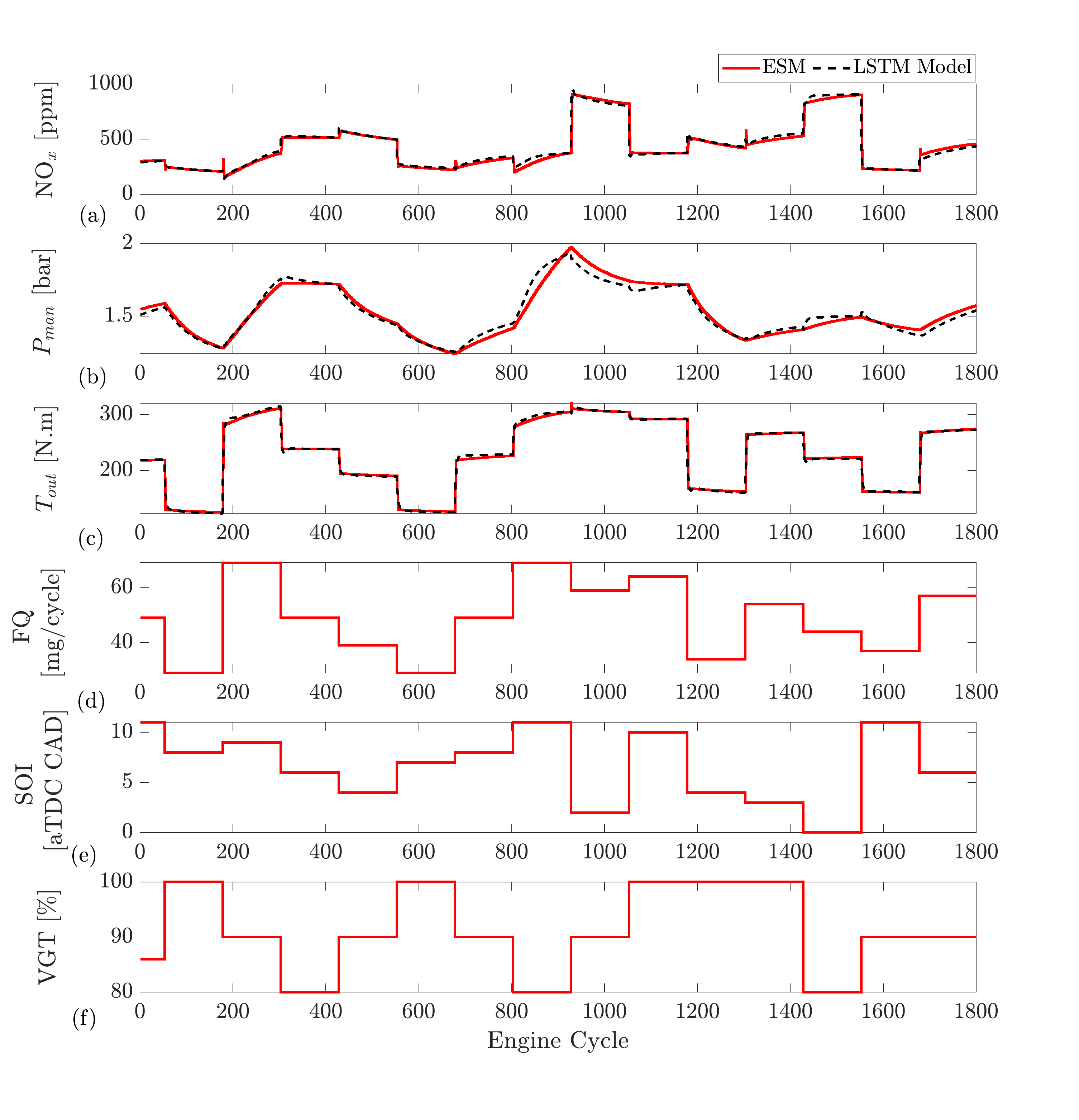}
    \caption{LSTM model comparison for engine-out emissions and performance: a) engine-out \nox, b) intake manifold pressure ($P_{\text{man}}$), c) engine-output torque ($T_{\text{out}}$), d) fuel quantity ($\text{FQ}$), e) Start of injection ($\text{SOI}$), f) Variable Geometry Turbine ($\text{VGT}$) rate}
    \label{fig:comparisonmodel}
\end{figure}

As this developed model will be used inside a Nonlinear Model Predictive Controller (NMPC) the accuracy of these models is critical. To compare the models, the LSTM model is run simultaneously in one simulation where the prediction of the LSTM models is compared against the ESM co-simulation. In this manner, the LSTM model is evaluated using newly generated test data that is new for the models. Figure~\ref{fig:comparisonmodel} shows the model comparison where the LSTM model is capable of estimating all outputs to a high accuracy. 

To summarize the accuracy of the model, the Normalized Root Means Square Error (NRMSE) between the ESM and LSTM model are calculated. The RMSE is normalized using the range (defined as the maximum value minus the minimum value) of the ESM logged data. The NRMSE for \nox, $P_{\text{man}}$, and $T_{\text{out}}$ are 4.04\%, 2.21\%, and 2.35\%, respectively. As shown, this model is capable of predicting emission and output torque with an error of less that 5\%.

\section{Nonlinear Model Predictive Controller Design}

Model predictive control (MPC) uses the principle of receding horizons in order to solve an optimization problem at each sampling instance to determine the control input. Here the objective of the controller is to minimize engine-out \nox~ emissions and fuel consumption while meeting the requested engine-output torque. The cost function $J (\bm{u}(\cdot|k), s(k))$ of the finite horizon optimal control problem (OCP) with horizon length $ p $ is defined as
\begin{equation} \label{eq:mpccostfunc}
\begin{split}
    J(\bm{u}(\cdot|k), s(k)) &= \sum_{i = 0}^{p-1} \Big[ \underbrace{||T_{\text{out}}(k+i) - T_{\text{out, ref}}(k+i) ||^2_{w_{T_{\text{out}}} }}_{\text{Torque output tracking}} \\
    &+ \underbrace{||\text{NO}_x(k+i)||^2_{w_{\text{NO}_x}}}_{\text{NO$_x$ minimization}} + \underbrace{||\text{FQ}(k+i)||^2_{w_{FQ}}}_{\text{fuel consumption minimization}}\\
    &+ \underbrace{||u(k+i|k) - u (k +i -1|k)||^2_{w_{\Delta u}}}_{\text{control effort penalty}} \\
    &+ \underbrace{w_{s} s(k)^2 }_{\text{Constraint violation penalty}} \Big]
\end{split}
\end{equation}
where 
\begin{equation}
    ||.||^2_w = [.]^Tw[.]
\end{equation}
and $s(k)$ is a nonnegative slack variable used to penalize the MPC cost function in the worst-case constraint violation, $w_{v}$, $ v \in [T_{\text{out}}, \text{NO}_x, \text{FQ}, \Delta u, s ]$, are the MPC weights, and $z_k^T$ is the optimization decision defined as 
\begin{equation}
    \bm{u}(\cdot|k) = [ u(k|k)^T \ \   u(k+1|k)^T \ \ ... \ \ u(k+p-1|k)^T]
\end{equation}

The model's outputs and manipulated variables are defined as
\begin{align}
\begin{split}
    y(k) &= \begin{bmatrix} T_\text{out}(k) & \text{NO}_x(k) \end{bmatrix}^T, 
    \\
	u(k) &= \begin{bmatrix} \text{FQ}(k) & \text{SOI}(k) & \text{VGT}(k) \end{bmatrix}^T.
\end{split}\label{eq:statescontrols}
\end{align}

In the defined cost function (Eq.~\ref{eq:mpccostfunc}) the error between output and requested torque are penalized along with \nox~ and fuel quantity minimization.

This leads to the definition of the following OCP, which is solved at each discrete-time instant:
\begin{equation}\label{eq:OCP}
\begin{aligned}
\min_{\bm{u}(\cdot|k), s(k)} \quad & J(\bm{u}(\cdot|k), s(k))\\
\textrm{s.t.} \quad & x(0) = \bar{x}(0) \\ 
\quad & x({k+1})= f(x(k), u(k)) \quad k = 0, \ldots, p-1 \\
\quad & \underline{x}\leq x(k) \leq \overline{x} \quad \quad\quad\quad\quad~ k = 0, \ldots, p \\
\quad & \underline{u}\leq u(k) \leq \overline{u} \quad \quad\quad\quad\quad~ k = 0, \ldots, p-1 \\
\end{aligned}
\end{equation}

\begin{figure}[h!]
    \centering
    \includegraphics[width = 0.49\textwidth]{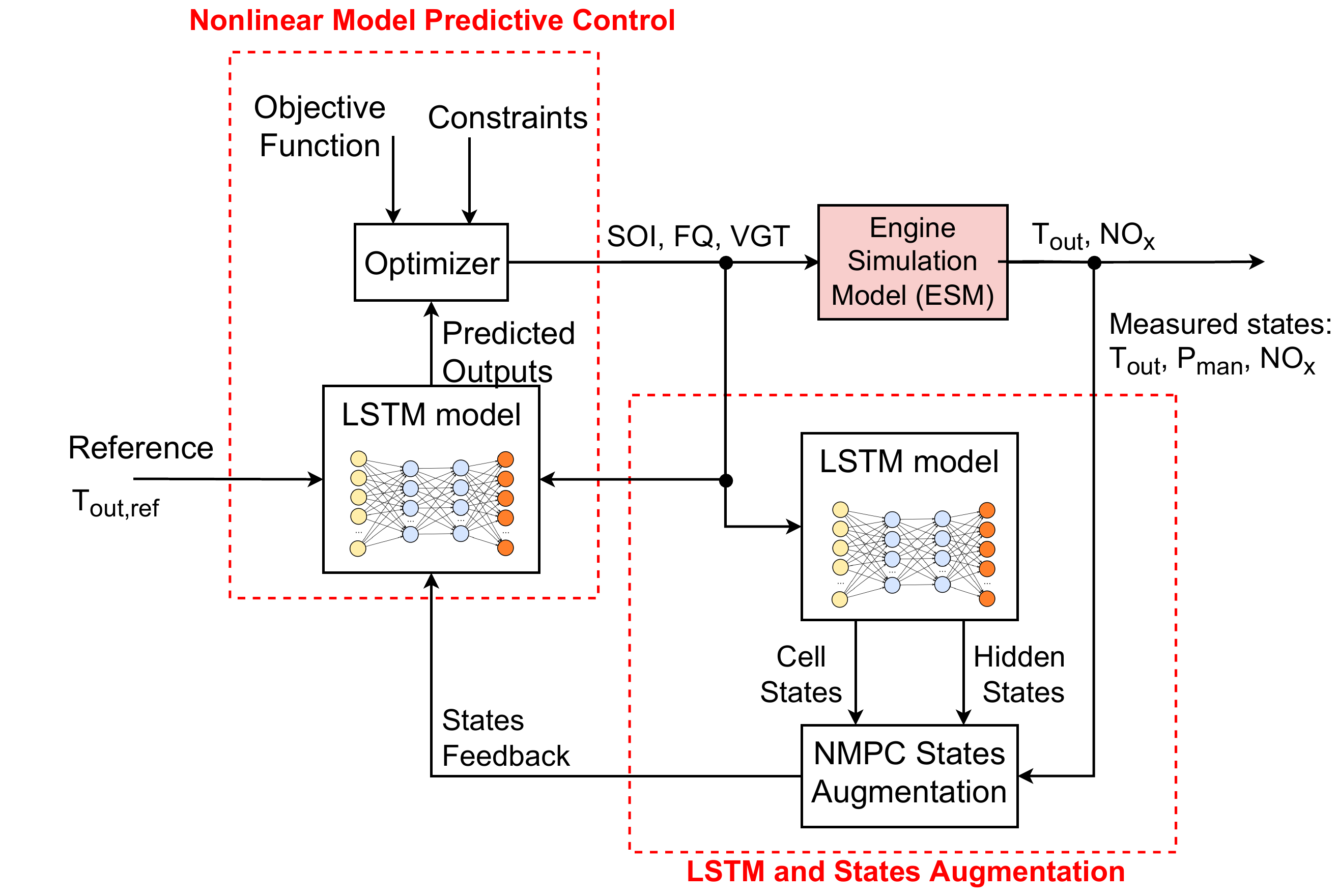}
    \caption{Block diagram of LSTM-NMPC structure}
    \label{fig:lstmmpc}
\end{figure}

Where $x_{k+1}= f(x(k), u(k))$ is from the LSTM model where $x$ represents the states of model. For the formulation of the LSTM-NMPC, the state vector of the MPC (Eq. \ref{eq:statescontrols}) has to be extended by adding the hidden and cell states of the LSTM network. Therefore, the prediction model of the LSTM-NMPC consists of the following updated state vector $ x(k)$, the output vector $ y(k) $ and the control vector $ u(k) $ which are defined as
\begin{align}
	{x} &= \begin{bmatrix} h_{LSTM1}, h_{LSTM2}
    c_{LSTM1}, c_{LSTM2} \end{bmatrix}^T , \notag \\
    {y} &= \begin{bmatrix} T_{\text{out}}, \text{NO}_x \end{bmatrix}^T \notag \\
	{u} &= \begin{bmatrix} \text{FQ}, \text{SOI}, \text{VGT} \end{bmatrix}^T,\label{eq:statescontrols_lstm}\notag .
\end{align}
The hidden and cell states result in a total of 104 states for the LSTM-MPC. Schematically the LSTM-NMPC is shown in Figure~\ref{fig:lstmmpc}. The nonlinear dynamics of the NMPC are defined by formulating the chain rule on the recurrent network´s prediction and update functions, stated in Eq.~\ref{eq:LSTMeq} and Eq.~\ref{eq:z_FC}. The prediction horizon, $p$, has been chosen as 5 with a control horizon of one step.

The NMPC problem has 104 states, and all of these states must be estimated to update the NMPC block that is used in MATLAB/SIMULINK\(^{\copyright}\). Therefore, the same dynamic model is used as an estimator for the hidden and cell states of the LSTM model to provide the 104 states of the NMPC problem.

The upper and lower constraints of the states and control output are given in Table~\ref{tab:mpc_constraints}. The upper limit on \nox~ is used to limit the peak emissions values and this value can be adjusted to comply with emissions legislation. For this work, a limit of 500ppm has been selected which is the maximum measured \nox emissions at 1500~rpm when using the production ECU.

The limit imposed on FQ is not strictly required as the goal of the MPC is to minimize FQ. However, the limit is imposed to prevent the controller from attempting large fuel injection amounts. Small injection amounts outside of the injector calibration range are also restricted. There is both an upper and lower constraint applied to the SOI. Early SOI is limited to prevent extremely early combustion phasing which can lead to high engine noise and possible engine damage. Late SOI is also constrained to prevent low combustion efficiency and high exhaust gas temperatures. The constraints provided on VGT are used to match the defined turbine map which runs from 0.7 - 1.0.

\begin{table} 
\begin{center}
\caption{Constraint Values}\label{tab:mpc_constraints}
\begin{tabular}{ccc}
\hline
Min Value ($\underline{x}, \underline{u})$ & Variable $(x,u)$ & Max Value $(\overline{x}, \overline{u})$ \\ \hline
$0\ \mathrm{ppm}$ & $NO\ind{x}$ & $500\ \mathrm{ppm}$ \\
$10\ \mathrm{mg/cycle}$ & $\text{FQ}$ & $80\ \mathrm{mg/cycle}$ \\
$-2\ \mathrm{aTDC~CAD}$& $\text{SOI}$ & $11\ \mathrm{aTDC~CAD}$ \\
$70\ \mathrm{\%}$& $\text{VGT}$ & $100\ \mathrm{\%}$\\
\hline
\end{tabular}
\end{center}
\end{table}

The weights of LSTM-NMPC are manually tuned where the highest weight is given to output torque tracking and giving equal weights to \nox and PM emissions and fuel consumption reduction. Then a weight equal to half that of the emissions is given to changes in control output. These tuned weights are $w_{T_{\text{out}}} = 1$, $w_{\text{NO}_x} = 0.2$, $w_{FQ} = 0.2 $, $w_{\Delta u} = 0.1$, respectfully. 

Here, the states, outputs, and control inputs are normalized using z-score normalization which is defined as 
\begin{equation}
    z_{\text{normal}} = \frac{z - \mu}{\sigma} 
\end{equation}
where $\mu$ is the mean and $\sigma$ is the standard deviation of $z$ for both the input and output space. This allows for a reduction in the computational time for the normalization and denormalization required for evaluating the LSTM neural network in the NMPC problem. The constraint softening value was set as 0.1 which indicates hard constraints. 

In this study, the MATLAB Toolbox utilizing the \texttt{fmincon} solver is used for the NMPC simulation. For comparison, state-of-the-art commercial solver \texttt{FORCES PRO} by EMBOTECH \cite{FORCESPro, FORCESNLP} and open-source package \texttt{acados}~\cite{Verschueren2021, Verschueren2018} with the QP solver \texttt{HPIPM} (High-Performance Interior-Point Method) \cite{frison2020hpipm} are also used to solve the NMPC optimization. In NMPC, the OCP structured NLP is reformulated into a Sequential Quadratic Programming (SQP) problem by means of iterative quadratic approximations at the shooting nodes~\cite{WINKLER2021359}. This results in sequential quadratic subproblems that can be solved and contribute to the holistic solution of the NLP by means of reformulations of the initial problem. The number of SQP iterations is called SQP steps which have a strong influence on the computational efficiency as well as the prediction quality of the NMPC. In the results and discussion section, the computational timing of \texttt{fmincon}, \texttt{FORCES PRO}, and \texttt{acados} with \texttt{HPIPM} are compared.

\section{NMPC Imitative Controller}

In addition to applying ML to modeling of the system it is also possible to utilize ML for control implementation. The imitation of NMPC using deep learning methods in so called imitative NMPC, is used to avoid the high computational time of online optimized NMPC~\cite{zhang2019safe}. 

\begin{figure}[h!]
    \centering
    \includegraphics[trim = 0 0 0 0, clip, width = 0.49\textwidth]{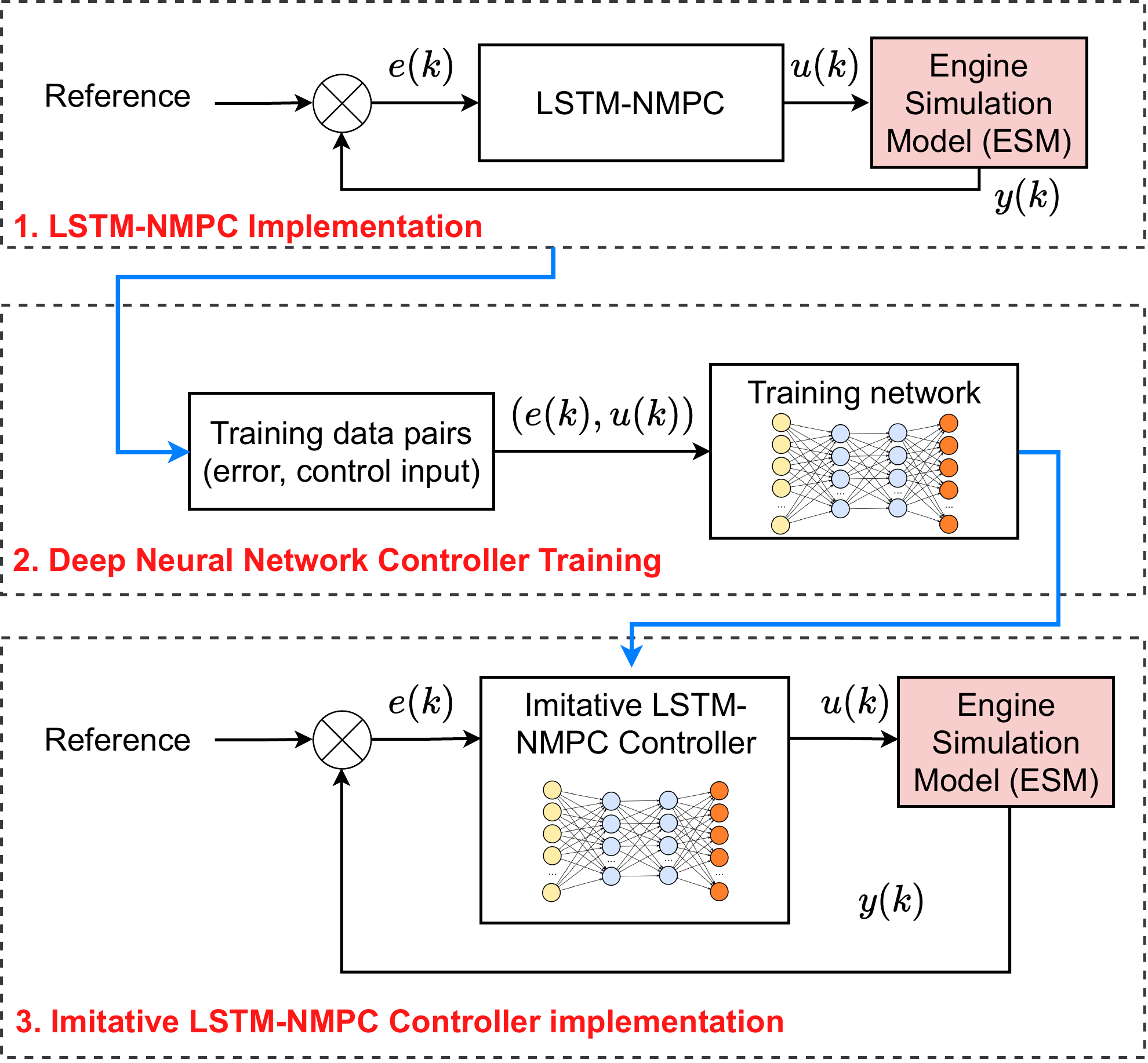}
    \caption{Concept of Imitative NMPC-- 1) Implementation of original LSTM-NMPC in ESM co-simulation, 2) Training deep neural network based on LSTM-NMPC collected data, 3) replace trained deep network (imitative controller) with LSTM-NMPC in ESM co-simulation-- NMPC: Nonlinear Model Predictive Contrl, LSTM: Long-Short Term Memory, ESM: Engine Simulation Model}
    \label{fig:imitaitvempcconcept}
\end{figure}

The application of imitative NMPC is shown schematically in Figure~\ref{fig:imitaitvempcconcept} which shown the three main steps. In the first step, the previously designed LSTM-NMPC is implemented on the ESM. Second, the NMPC input and output are recorded, and a deep neural network, including a LSTM layer is used to fit the controller data to mimic the behavior of the NMPC. The final step in the process involves replacing the online NMPC with an imitative controller to avoid the high computational time of NMPC. That is, instead of solving NMPC optimization online, the identified function, here a deep network, is deployed with a much lower computation cost. 

\begin{figure}[h!]
    \centering
    \includegraphics[trim = 0 0 0 0, clip, width = 0.49\textwidth]{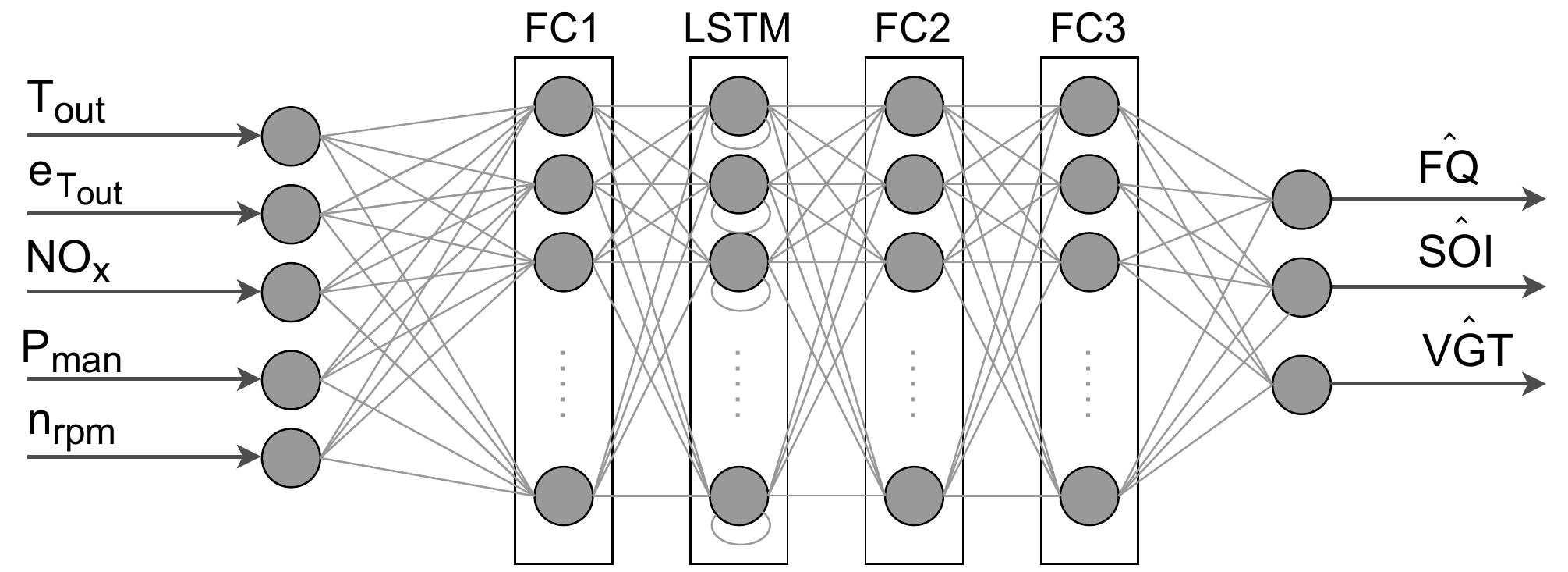}
    \caption{Structure of proposed network for imitation of NMPC}
    \label{fig:StructureOfimitative}
\end{figure}

In order to clone the behavior of the NMPC, a deep network structure using LSTM is proposed. The proposed imitative LSTM-NMPC is shown schematically in Figure~\ref{fig:StructureOfimitative}. The engine-output torque \( T_{\text{out}}\), the error in output torque (\( e_{T_{\text{out}}}\)), engine-out \nox, intake manifold pressure \( P_{\text{man}} \), and engine speed \( n_{rpm}\) are the inputs of the imitative NMPC, and the goal here is to generate control action of fuel quantity \( (\text{FQ}) \), start of injection \( (\text{SOI}) \), and VGT by mimicking the previously designed NMPC controller.

Therefore, the forward propagation of this imitative network used to replace the online optimization of MPC is given as
\begin{subequations}
\begin{align}
\begin{split}
z_{FC1}(k) &= \text{ReLU}\left(W_{FC1}^T {u(k)} + b_{FC1}\right)\\
\end{split}\\
\begin{split}
h_{LSTM}{(k)} &= \underbrace{o_{LSTM}{(k)} \odot \text{tanh}\left(c_{LSTM}{(k)}\right)}_{\text{based on Eq.}~\ref{eq:LSTMeq}}\\
\end{split}\\
\begin{split}
z_{FC2}(k) &= \text{ReLU}\left(W_{FC2}^T h_{LSTM}{(k)} + b_{FC2} \right)\\
\end{split}\\
\begin{split}
\underbrace{z_{FC3}(k)}_{\hat{u}(k) = [\hat{\text{FQ}}(k) \ \ \hat{\text{SOI}}(k) \ \ \hat{\text{VGT}}]^T}  &= \text{ReLU}\left(W_{FC3}^T z_{FC2}(k) + b_{FC3}\right)\\
\end{split}
\end{align}
\end{subequations}

The output of the proposed imitative LSTM-NMPC network which is the estimated control actions can be calculated as
\begin{equation}
    \begin{split}
    \hat{u}(k) &= \left[\begin{array}{c}
  \hat{\text{FQ}}(k) \\
   \hat{\text{SOI}}(k) \\
   \hat{\text{VGT}}\\
  \end{array} \right]
    \end{split}
\end{equation}
The cost function for this network is
\begin{equation}\label{eq:costreg1}
J(W, b) = \frac{1}{m}  \sum_{k = 1}^m \mathcal{L}\left(\hat{u}(k), u(k)\right) + \frac{\lambda}{2m} \sum_{l = 1}^L ||W^{[l]}||_2^2
\end{equation}
where \(\mathcal{L}\left(\hat{u}(k), u(k)\right)\) is the loss function, \(\lambda\) is the regularization coefficient, and \(||W^{[l]}||_2^2\) is the Euclidean norm which is defined as
\begin{equation}\label{eq:wreg1}
||W^{[l]}||_2^2 = \sum_{i = 1}^{n^{[l]}} \sum_{j = 1}^{n^{[l-1]}} (w_{ij}^{[l]})^2
\end{equation}
The Mean Squared Error (MSE) cost function is used and is defined as
\begin{equation}
    \mathcal{L}\left(\hat{u}(k), u(k)\right) = \frac{1}{m} \sum_{k = 1}^m (\hat{u}(k) - u(k))^2
\end{equation}

\begin{figure}
     \centering
    \includegraphics[trim = 0 0 0 0, clip, width = 0.49\textwidth]{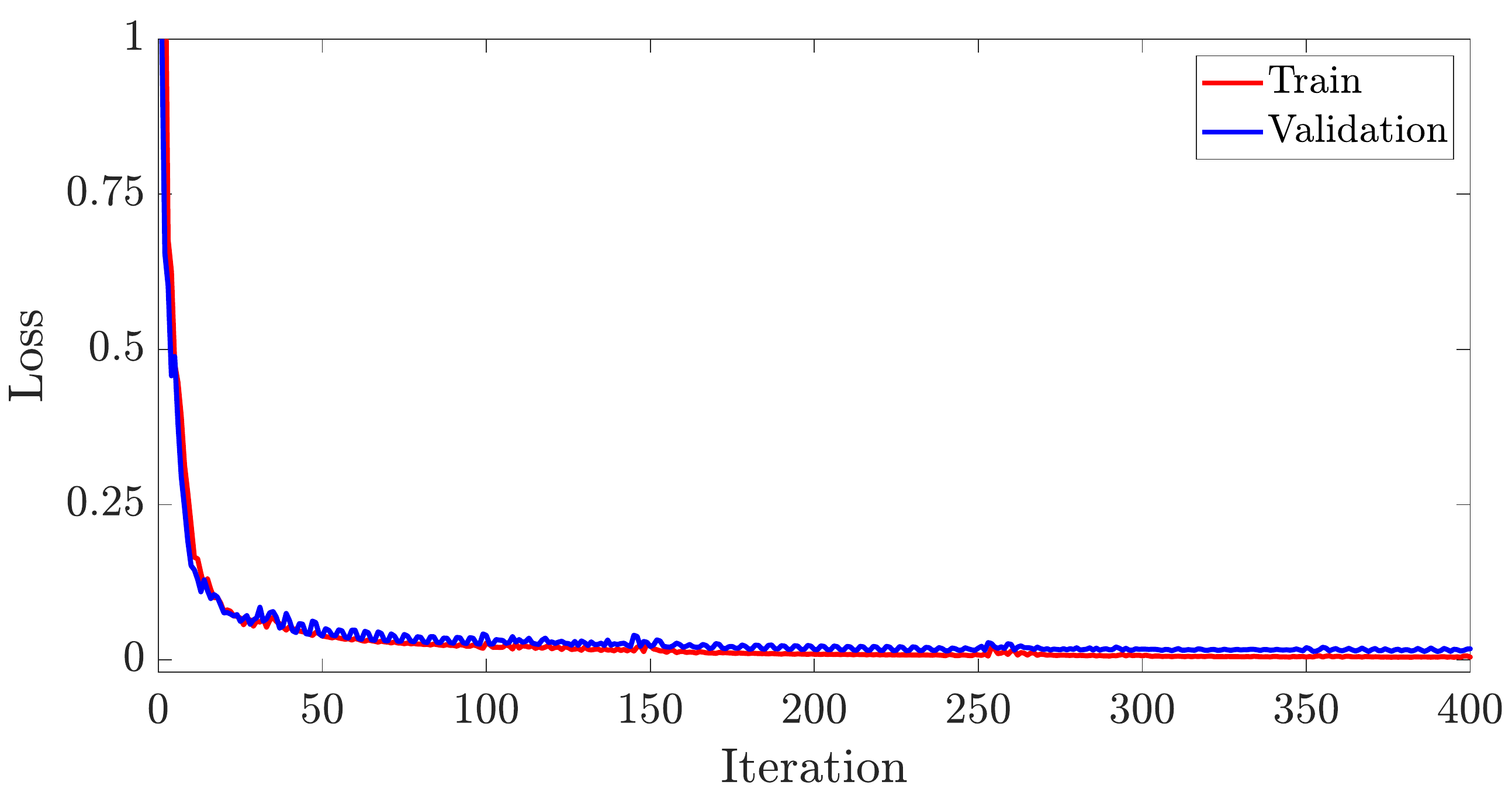}
        \caption{Imitative LSTM-NMPC loss function vs. iteration}
    \label{fig:imitativeloss}
\end{figure}

The training details of the imitative controller is presented in Table~\ref{tab:imitativempc}. As the performance of the imitative controller depends on collected data, widely varying operating conditions need to be tested and fed into the network. To do this the NMPC controller was evaluated for randomly changing engine speeds from 1200 rpm to 1800 rpm and for a requested load ( \( T_{\text{out}}\) reference) of 120 - 320 N.m to make the imitative controller robust to a range of operating conditions. The loss function versus iteration is shown in Figure~\ref{fig:imitativeloss}. This loss function indicates the training process has been completed and has avoided the overfitting-underfitting problem as the validation data eventually converges to training loss. 

\begin{table}[h!]
    \centering
    \caption{Properties of imitative controller based on LSTM-NMPC}
    \begin{tabular}{l| c c}
    \hline
       \textbf{Name}  &  \textbf{LSTM-NMPC }\\
       \hline
        FC size & 32\\
        LSTM size &32\\
        Optimizer & Adam\\
        Maximum Epochs& 400\\
        Mini batch size & 512\\
        Learn rate drop period& 150 Epochs\\
        Learn rate drop factor & 0.5\\
        L2 Regularization& 1 \\
        Initial learning rate & 0.02\\
        Validation frequency & 1\\
        Momentum & 0.9\\
        Squared gradient decay & 0.99\\
        \hline
    \end{tabular}
    \label{tab:imitativempc}
\end{table}

To train the imitative controllers, the NMPC is evaluated for 2000 seconds of simulation, where 1600 seconds are devoted for training data and 400 seconds of the simulation are used as the validation data set. The RMSE of the training and validation set for are presented in Table.~\ref{tab:imitativecontrolleraccuracy}. As shown, the DNN network can clone the NMPC behavior with average accuracy of 3.9\% for training and 8.3\% for validation. These imitative controllers are tested for newly generated references, and their performance against NMPC online optimization will be presented in the next section.

\begin{table}[h!]
    \centering
     \caption{Imitative LSTM-NMPC controller train and validation accuracy}
    \begin{tabular}{c | c c c}
    \hline
        & \textbf{\text{FQ}}  & \textbf{SOI} & \textbf{VGT} \\
    \hline
    \textbf{Train} & 2.12\% & 4.55\% & 4.90\% \\
        \textbf{Validation} & 2.47\% & 9.98\%  & 12.46\%\\
        \hline
    \end{tabular}
    \label{tab:imitativecontrolleraccuracy}
\end{table}

\section{Results and Discussions}

\begin{figure}[ht!]
     \centering
    \includegraphics[trim = 10 50 70 0, clip, width = 0.49\textwidth]{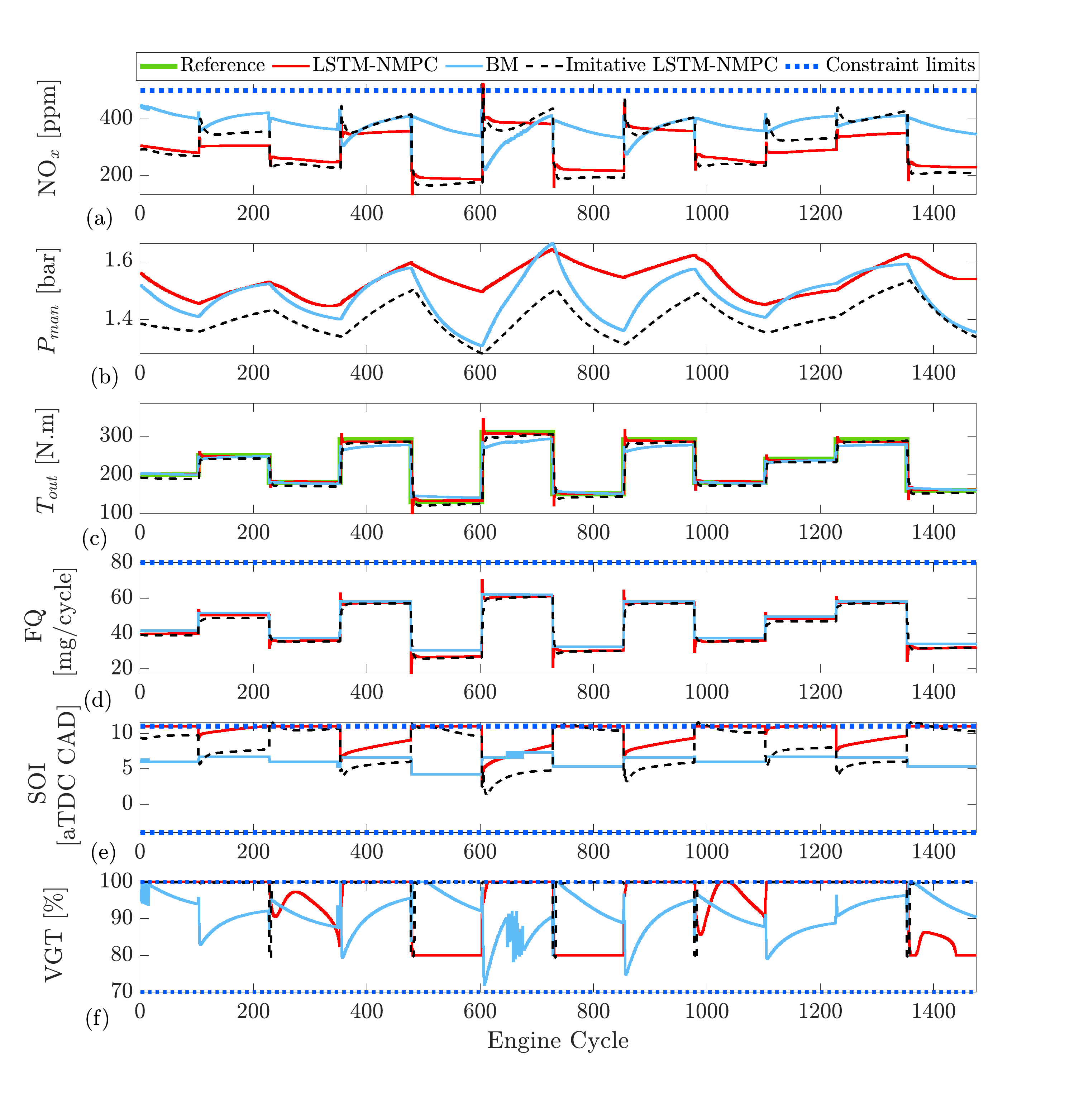}
        \caption{Controller comparison in engine speed of 1500 rpm: a) engine-out \nox, b) intake manifold pressure ($P_{\text{man}}$), c) engine-output torque ($T_{\text{out}}$), d) fuel quantity ($\text{FQ}$), e) Start of injection ($\text{SOI}$), f) Variable Geometry Turbine ($\text{VGT}$) rate}
    \label{fig:com1500}
\end{figure}

The objective of the developed controller is to track a target load while minimizing \nox~and the fuel used in the diesel engine. Here the results of the NMPC controller with LSTM-based data driven model and imitative LSTM-NMPC controller are compared to the benchmark model (BM). The BM here is the calibrated ECU tables based on the Cummins production ECU that is embedded in GT-power for ESM co-simulation-- more details are available in our previous studies~\cite{saeed2021MECC, shahpournyenergies}. Figure~\ref{fig:com1500} presents all the controller inputs and outputs for all three controllers tested.

{The tested controllers are subject to input constraints for \nox, FQ, SOI and VGT.} These limits are listed in Table~\ref{tab:mpc_constraints}. Slight constraint violations for \nox~ at engine cycle 600 are seen. This is attributed to the constraint softening that is implemented in the NMPC. However, overall both the LSTM-NMPC and imitative controllers are able to keep the \nox~ levels below the specified constraint. The BM is significantly worse than the nonlinear MPC model which is likely due to operating the engine not at one of the optimal calibration points in the feedforward tables. 

The upper limit for SOI constraints is often fhit. Here the controllers would like to implement later injection timing than is currently allowed. These late injection timings reduce the peak combustion temperature leading to lower \nox~ levels, however, they also lead to reduced combustion efficiency and increased fuel consumption.

\begin{figure}[ht!]
     \centering
    \includegraphics[trim = 10 50 70 0, clip, width = 0.49\textwidth]{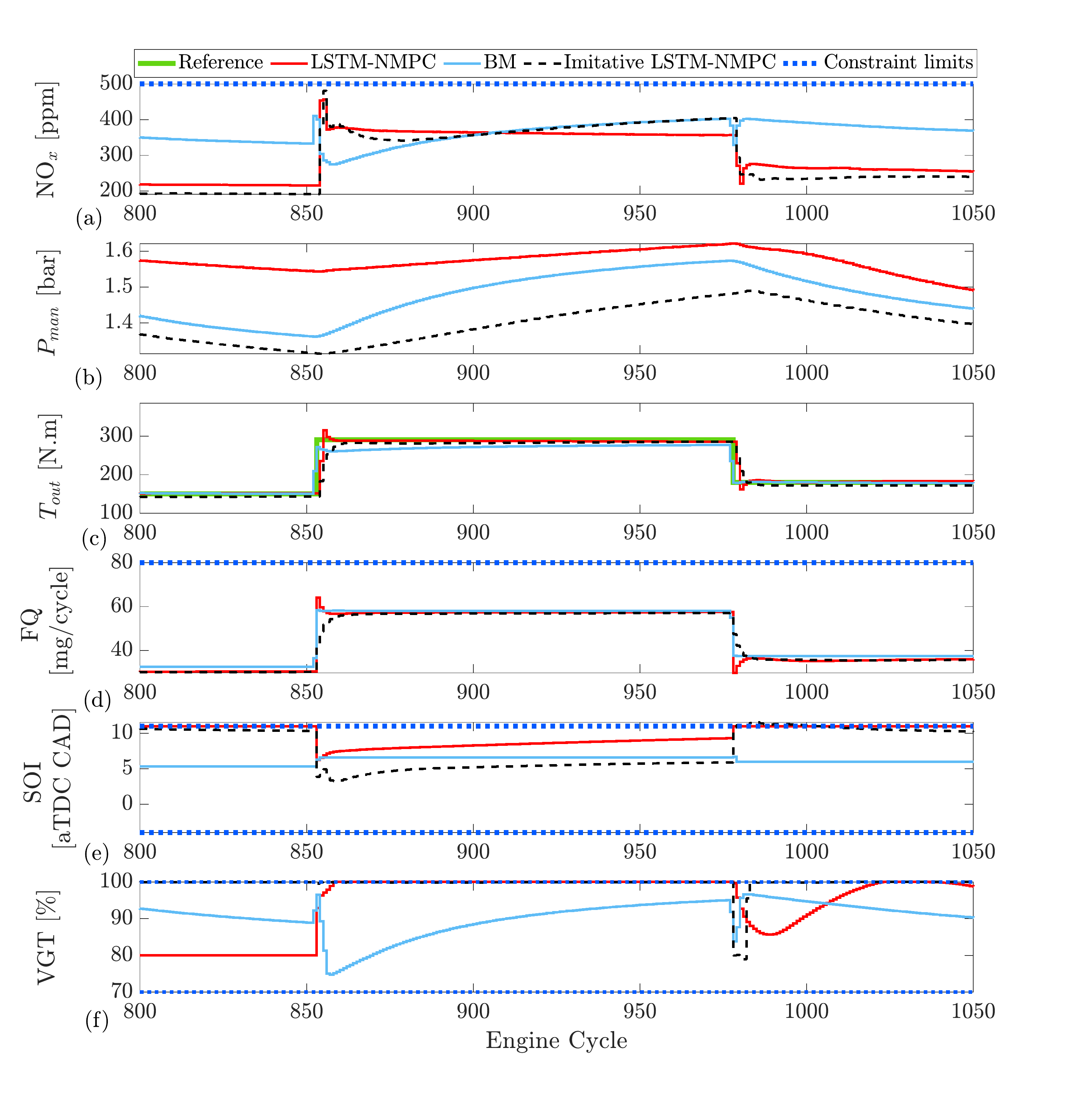}
        \caption{Controller comparison in engine speed of 1500 rpm zoomed from 800 to 1050 cycles: a) engine-out \nox, b) intake manifold pressure ($P_{\text{man}}$), c) engine-output torque ($T_{\text{out}}$), d) fuel quantity ($\text{FQ}$), e) Start of injection ($\text{SOI}$), f) Variable Geometry Turbine ($\text{VGT}$) rate}
    \label{fig:com1500z}
\end{figure}

The torque ($T_{\text{out}}$) tracking of both NMPC and BM are shown for a step in Figure~\ref{fig:com1500z} and acceptable performance is achieved. Here the BM uses the most fuel and still is unable to reach the desired torque.

\begin{figure}[ht!]
     \centering
    \includegraphics[trim = 10 50 70 0, clip, width = 0.49\textwidth]{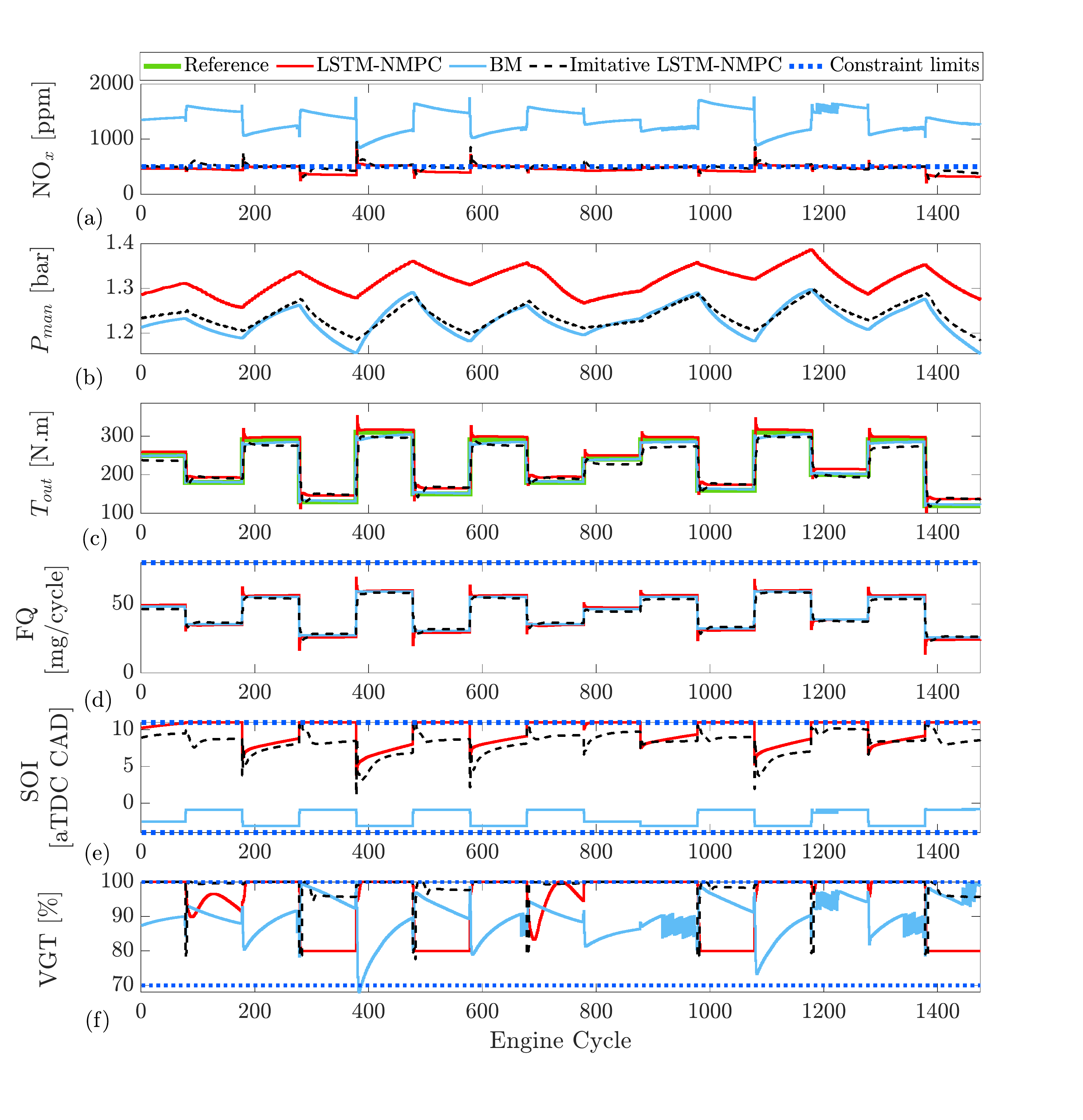}
        \caption{Controller comparison in engine speed of 1200 rpm: a) engine-out \nox, b) intake manifold pressure ($P_{\text{man}}$), c) engine-output torque ($T_{\text{out}}$), d) fuel quantity ($\text{FQ}$), e) Start of injection ($\text{SOI}$), f) Variable Geometry Turbine ($\text{VGT}$) rate}
    \label{fig:com1200}
\end{figure}

\begin{figure}[ht!]
     \centering
    \includegraphics[trim = 10 50 70 0, clip, width = 0.49\textwidth]{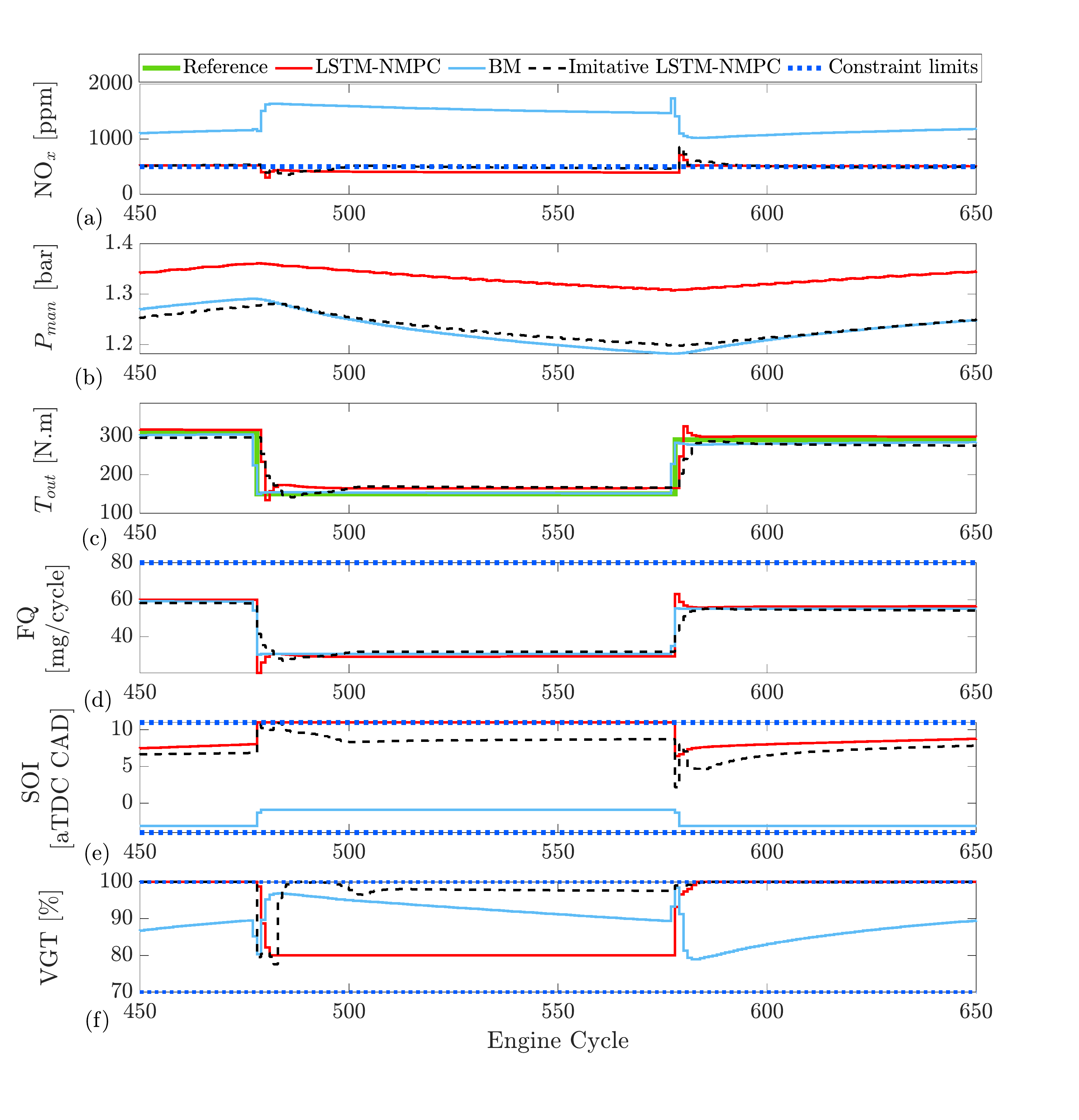}
        \caption{Controller comparison in engine speed of 1200 rpm zoomed from 450 to 650 cycles: a) engine-out \nox, b) intake manifold pressure ($P_{\text{man}}$), c) engine-output torque ($T_{\text{out}}$), d) fuel quantity ($\text{FQ}$), e) Start of injection ($\text{SOI}$), f) Variable Geometry Turbine ($\text{VGT}$) rate}
    \label{fig:com1200z}
\end{figure}

To check the controller's robustness, the engine speed is changed from 1500~rpm (where the NMPC model was designed and validated) to 1200~rpm. The NMPC performance at 1200~rpm can be seen in Figure~\ref{fig:com1200} (zoomed version from 450 to 650 is shown in Figure~\ref{fig:com1200z}). {From these figures it can be seen that both the LSTM-NMPC and Imitative NMPC exhibit similar trends to their operation at 1500~rpm. Again the actions of the imitative controller are slower and without overshoot to the step inputs when compared to the online LSTM-NMPC. Both of these controllers outperform the BM production ECU in terms of \nox production. As the speed is changed from 1500~rpm, the imitative controller has oscillations in the controller outputs as seen in fuel quantity, start of injection and VGT rate. The largest effect of changing speed is the significant increase in \nox emissions of the BM. It is assumed that this is caused by deviating from 1500~rpm, the design operating point of the industrial engine. The high \nox~value at this speed in the BM is due to a significant advance in injection timing resulting in a significant increase in \nox~ emissions. }

To summarize the performance of the controllers at both engine speeds the cumulative and average \nox~ emissions, load error, FQ and Execution time are shown in Table~\ref{tab:controllerresults}. For the design speed of 1500~rpm both the LSTM-NMPC and imitative NMPC outperform the BM in terms of \nox~ emission with a slight reduction in fuel consumption. 

To investigate the NMPC execution time, \texttt{acados} (QP solver \texttt{HPIPM}), \texttt{FORCES PRO}, and \texttt{fmincon} are each evaluated and their computational time are compared and summarized in Table~\ref{tab:turnaround}. \texttt{acados} with \texttt{HPIPM} provides the fastest solve time among the solvers tested. In the \texttt{acados} implementation, a maximum QP iteration of 50 and maximum NLP iteration / SQP steps of 5 are used, which results in an average runtime of 12.20 ms and maximum runtime of 31.56 ms for 1500 RPM. This value is much faster than than the average runtimes of \texttt{fmincon} that required 786.02 ms. \texttt{FORCES PRO} was also tested and and showed an improvement in runtime over the \texttt{fmincon} but was was significantly slower than the \texttt{acados} implementation. The exact computational timing of \texttt{FORCES PRO} cannot be disclosed due to the academic license agreement.

\begin{table*}
    \centering
    \begin{tabular}{l|cc}
    \hline
       \textbf{Solver}  &  \textbf{Average Turnaround time [ms]} & \textbf{Real-time verification in PIL setup}\\
    \hline
        Matlab \texttt{fmincon} &  786.02 & x \\
        EMBOTECH \texttt{FORCES PRO} &  786.02 $>$ t $>>$ 12.20 & x \\
        \texttt{acados} & 12.20 &  \checkmark \\
        \hline
    \end{tabular}
    \caption{Turnaround time comparison between Matlab \texttt{fmincon}, EMBOTECH \texttt{FORCES PRO}, and \texttt{acados} solvers-- PIL: Processor in the loop}
    \label{tab:turnaround}
\end{table*}

The authors assume that this difference in the runtime can be attributed to the fully condensed problem formulation used in the \texttt{acados} implementation in comparison to the sparse formulation of \texttt{fmincon} and \texttt{FORCES PRO}. The presented OCP consists of a large state vector with 107 states and a relatively small prediction horizon with 5 steps and a relatively small control vector with 3 control outputs. This allows the condensed problem formulation used in \texttt{acados} to take full advantage of the condensation benefits~\cite{Diehl2009, Frison2016}. Another cause for the difference in runtime is attributed to the underlying algorithmic differentiation framework \texttt{CasADi} MX (Matrix expression) symbolic variables in \texttt{acados} instead of SX (scalar symbolic) variables in \texttt{FORCES PRO}. The general matrix expression type MX tends to be more computationally efficient when working with larger matrices~\cite{Andersson2019}.

Although the \texttt{acados} computational time of the online NPMC is orders of magnitude slower than imitation NMPC. The ML-based imitation controllers rely on MPC results, and thus an MPC must be designed and run to collect data for controller training. {The ECU time of the benchmark model is estimated by recreating the production feedforward calibration tables on the MABX~II. As there are various corrections applied to the calculated values, the BM controller actually takes longer to compute than the imitative controller. This is expected as the BM model considers various environmental factors such as engine and air temperatures as well as ambient pressures, which are not considered in the developed NMPC.   }

\begin{table*}[ht!]
    \centering
    \caption{Proposed MPC and Imitative MPC results compared to Benchmark for engine speed of 1500 and 1200 rpm}
    \begin{tabular}{l | c c c c c c}
        \hline
        \hline
        \multicolumn{7}{c}{\textbf{1500 rpm}}\\
        \hline
        \hline
        &Cumulative & Average & Load & Cumulative & Average  & Execution \\
        &\nox~ [ppm] & \nox~ [ppm] &error [\%] & \text{FQ} [g] &  \text{FQ} [mg]  & time [ms]$^*$ \\
        \hline
        Benchmark & 556100.0 & 376.8 & 3.95 & 67.9 & 46.0 & $\approx$0.08$^{**}$ \\
        LSTM-NMPC & 428420.0 & 290.2 & 1.90 & 65.6 & 44.4 & 12.20$^{***}$ \\
        Imitative LSTM-NMPC & 438850.0 & 297.3 & 3.74 & 64.6 & 43.8 & 0.04 \\
        \hline
        \hline
        \multicolumn{7}{c}{\textbf{1200 rpm}}\\
        \hline
        \hline
        &Cumulative & Average & Load & Cumulative & Average  & Execution \\
        &\nox~ [ppm] & \nox~ [ppm] &error [\%] & \text{FQ} [g] &  \text{FQ} [mg]  & time [s]$^*$ \\
        \hline
        Benchmark & 1961900.0 & 1329.2 & 2.18 & 64.8 & 43.9 & $\approx$0.08$^{**}$\\
        LSTM-NMPC & 671650.0 & 455.0 & 6.95 & 64.9 & 43.9 & 11.50$^{**}$\\
        Imitative LSTM-NMPC & 718380.0 & 486.7 & 7.61 & 64.0 & 43.4 & 0.03\\
        \hline
        \multicolumn{7}{l}{$^*$ { \small per engine cycle of simulation}}\\
        \multicolumn{7}{l}{$^{**}$ { \small Exact production ECU time is unknown, estimated by recreating Cummins ECU on MABX II}}\\
        \multicolumn{7}{l}{$^{***}$ { \small \texttt{acados} execution time}}\\
    \end{tabular}
    \label{tab:controllerresults}
\end{table*}

The performance of the developed controllers compared to the BM can be seen in Table~\ref{tab:savingcontroller}. Here significant \nox~ emissions reduction can be seen for the controller over the BM. In addition to these improvements in emissions the developed controllers use the same amount or less fuel than the baseline model. This demonstrates the advantage of the optimized controllers over the BM. The developed models do experience a slight increased load error in comparison to BM for 1500~rpm. However, the 2\% worse load tracking results in significant emission and fuel consumption benefits.

\begin{table*}[ht!]
    \centering
    \caption{Percentage of improvement for proposed MPC and Imitative MPC with respect to Benchmark for engine speed of 1500 and 1200 rpm}
    \begin{tabular}{l | c c c}
        \hline
        \hline
        \multicolumn{4}{c}{\textbf{1500 rpm}}\\
        \hline
        \hline
         & \nox~ [\%]  &  \text{FQ} [\%] & load error [\%]  \\
        \hline
        LSTM-NMPC & -22.98 & -3.48 & +2.05 \\
        Imitative LSTM-NMPC & -21.10 & -4.78 & +0.21  \\
        \hline
        \hline
        \multicolumn{4}{c}{\textbf{1200 rpm}}\\
        \hline
        \hline
        & \nox~ [\%]  & \text{FQ} [\%] & load error [\%]  \\
        \hline
        LSTM-NMPC & -65.77 & +0.15 & -4.77 \\
        Imitative LSTM-NMPC  & -63.38 & -1.23 & -5.43\\
        \hline
    \end{tabular}
    \label{tab:savingcontroller}
\end{table*}

{The developed controller in this study has also been compared to Linear Parameter Varying MPC (LPV-MPC) designed for ICE control in the literature~\cite{ICEMPC7}. The LPV-MPC adjustment for the emission control of a CI engine was developed in our previous study~\cite{norouzilpvmpc2022conf} based on~\cite{ICEMPC7}-- more details of this implementation are available in~\cite{norouzilpvmpc2022conf}. Table~\ref{tab:controllerresults_lpv} shows the performance of LPV-MPC vs. LSTM-NMPC and imitation LSTM-NMPC. As a result of using a nonlinear optimization (LSTM-NMPC), the computational time is increased by 10.5~ms when compared to LPV-MPC. However, the imitation of LSTM-NMPC provides more than 40 times faster performance than the LPV-MPC. Both LSTM-NMPC and imitation LSTM-NMPC are capable of reducing \nox~emission by 4.9\% and 2.6\% compared to LPV-NMPC. This \nox~reduction is a trade-off with a slight increase in the fuel consumption of LSTM-NMPC.}

\begin{table*}[ht!]
    \centering
    \caption{{Proposed MPC and Imitative MPC results compared to Linear Parameter Varying MPC (LPV-MPC) \cite{ICEMPC7, norouzilpvmpc2022conf} for engine speeds of 1500}}
    \begin{footnotesize}
    \begin{tabular}{l c c c c c c}
        \hline
        &Cumulative & Average & Load & Cumulative & Average  & Execution \\
        &NOx [ppm] & NOx [ppm] &error [\%] & FQ [g] &  FQ [mg]  & time [ms]\\
        \hline
        LPV-MPC~\cite{ICEMPC7, norouzilpvmpc2022conf} & 450570.0 & 305.3 & 2.96 & 65.5 & 44.4& 1.69 \\
        LSTM-NMPC & 428420.0 & 290.2 & 1.90 & 65.6 & 44.4 & 12.20 \\
        Imitative LSTM-NMPC & 438850.0 & 297.3 & 3.74 & 64.6 & 43.8 & 0.04 \\
        \hline
    \end{tabular}
    \end{footnotesize}
    \label{tab:controllerresults_lpv}
\end{table*}

{Overall, the imitative NMPC controller provide similar improvements to the NMPC controller over the BM and LPV-MPC~\cite{ICEMPC7, norouzilpvmpc2022conf} but provides a significant decrease in computational time. This makes future real-time implementation while minimizing the computational power needed with a full MPC feasible. }

\section{Conclusions}

The integration of machine learning and model predictive control for both modeling and controller implementation is discussed. First, using a deep recurrent neural network, a long-short term memory (LSTM) network is designed to predict diesel engine performance and emissions. A nonlinear model predictive controller (NMPC) is designed based on this network and by augmenting hidden and cell states. The prediction accuracy of these models for test data shows acceptable results. This model accuracy is expected as the LSTM is capable of a more generalizable prediction since it uses hidden and cell states to capture long term relationships within the data. 

In addition to using ML for modeling ML can also be implemented as a controller where an NMPC is implemented and data is collected from the MPC input and output which is used to train a deep neural network. By replacing the full online MPC by an ML based imitation it is possible to reduce the computational time of an NMPC. This imitative controller is compared with a baseline Cummins calibrated ECU model using the Engine Simulation Model (ESM) in a GT-power/MATLAB/SIMULINK co-simuation. 

Minimizing \nox~ emissions and reducing the injected fuel amount while maintaining the same load is the main goal of the controller. The developed controllers are also constrained to meet \nox legislated limits in addition to constraints on all inputs to ensure engine safety. The developed controllers are simulated using the ESM co-simulation. To evaluate the robustness of the controllers, the engine speed is changed from 1500~rpm where the NMPC model were validated to 1200~rpm. All of the controllers produce significant \nox~ reduction, especially at lower engine speeds with respect to benchmark feedforward controller. The \nox~ reduction for 1500 and 1200 rpm for NMPC are 23.0\% and 65.8\%. The imitative controller successfully cloned the NMPC behavior with a \nox~ reduction of 21.1\% at 1500 rpm. At 1200 rpm the reduction is 63.4\% when compared to the BM. The imitative controller performs similarly to the online MPC by learning from the MPC experiment but requires a much lower computational time. The computation time for the imitative controller is a factor of 100 lower than the online optimized MPC. Real-time implementation of LSTM-NMPC and imitative LSTM-NMPC controllers on the engine is planned as future work.

\section*{Acknowledgments}
The author(s) disclosed receipt of the following financial support for the research, authorship, and/or publication of this article: The research was performed as part of the Research Group (Forschungsgruppe) FOR 2401 “Optimization based Multiscale Control for Low Temperature Combustion Engines,” which is funded by the German Research Association (Deutsche Forschungsgemeinschaft, DFG) and with Natural Sciences Research Council of Canada Grant 2016-04646. Partial funding from Future Energy Systems and Alberta innovates at the University of Alberta is also gratefully acknowledged.

\section*{Declaration of competing interest}
The authors declare that they have no known competing financial interests or personal relationships that could have appeared to influence the work reported in this paper.

\bibliographystyle{unsrtabbv}

\bibliography{ML_ReviewAllrefs}

\end{document}